\documentclass[a4paper, 12pt]{article}

\usepackage[left=1.5cm, right=1.5cm, top=1.5cm]{geometry}
\usepackage{amsmath}
\usepackage{amssymb}
\usepackage{graphicx}
\usepackage{subcaption}
\usepackage{subfig}
\usepackage{dcolumn}
\usepackage{bm}

\usepackage[utf8]{inputenc}
\usepackage[T1]{fontenc}
\usepackage{mathptmx}
\usepackage{etoolbox}
\usepackage[square,numbers]{natbib}

\usepackage{enumitem}

\providecommand{\keywords}[1]{%
  \par\smallskip\noindent\textbf{Keywords:}~#1\par
}

\title{Raubold–Lynch construction of the phase space in hypertriton three-body mesonic decay}

\author{
Emile Meoto\\[0.6cm]
Department of Physics, University of Buea\\
P.~O.~Box 63, Buea, South West Region, Cameroon\\
Email: meoto.emile@ubuea.cm or emeotoson@gmail.com
}

\date{\today}

\begin{document}

\maketitle

\begin{abstract}
A Monte Carlo construction is presented for the three-body mesonic decay phase space of the hypertriton. This is accomplished using the Raubold–Lynch sequential two-body decay algorithm. The method factorises this three-body decay into successive two-body decays through a virtual intermediate subsystem, whose invariant mass is sampled over the full kinematically allowed region. Exact relativistic two-body kinematics and Lorentz transformations are employed to construct complete four-momenta for all decay products. The construction is applied to both charged- and neutral-pion decay channels, yielding events that reproduce the Lorentz-invariant three-body phase-space measure. The resulting momentum spectra, pairwise momentum correlations, opening-angle distributions, and Dalitz plots furnish a complete kinematic characterisation of both decay modes. The neutral-pion channel exhibits a larger available three-body energy and correspondingly broader kinematic limits. The unweighted events generated provide a kinematic baseline that can subsequently be weighted event by event with a weak-decay matrix element, for computing decay rates and other observables. For the neutral-pion channel, where no experimental data currently exists, the Lorentz-invariant phase-space constitutes a useful baseline for predictions.
\end{abstract}

\section{Introduction}

Following the discovery of the first hypernucleus in 1952, the hypertriton was identified as the lightest known hypernucleus and an ideal laboratory for investigating the lambda-nucleon interaction. In their pioneering study, Dalitz and Downs \cite{dal1958a} used the Q-values of mesonic decay events of the hypertriton  observed in experiments to determine its lambda separation energy. Through a variational three-body calculation of the hypertriton, they determined the strength and spin dependence of a phenomenological lambda-nucleon potential that reproduces the experimental lambda separation energy. Building directly on this work, Leon \cite{leo1959} (and many years later Kolesnikov and Kopylov\cite{kol1988}) carried out an accurate evaluation of the two-body mesonic branching fraction for the hypertriton. The subsequent work by Dalitz and Liu \cite{dal1959} provided the first dedicated theoretical treatment of the pionic decay modes of light lambda hypernuclei, including the hypertriton. The first quantitative calculation of the hypertriton mesonic decay rate and lifetime was performed by Rayet and Dalitz \cite{ray1966}. All of these pioneering studies focused on two-body mesonic decay. More recently, it was shown that the strong $\pi^-{}^3\mathrm{He}$ final-state interaction in the two-body decay ${}^3_{\Lambda}\mathrm{H} \rightarrow {}^3\mathrm{He}+\pi^-$ enhances the calculated decay width by approximately $18\%$ \cite{hil2023}. This result demonstrates that final-state interactions can produce significant dynamical modifications of the mesonic decay rate beyond the underlying kinematic phase space.

The three-body mesonic channels provide a natural extension in which the interplay between kinematics and dynamics becomes more involved. Negative-pion and neutral-pion decays are given by

\begin{align}
    {}^3_\Lambda\text{H} \rightarrow d + p + \pi^- \\
    {}^3_\Lambda\text{H} \rightarrow d + n + \pi^0
\end{align}

Experimentally, the three-body negative pion mesonic decay of the hypertriton has been directly measured by the STAR Collaboration in a relativistic heavy ion collider experiment \cite{ada2018}. Earlier measurements of this decay mode were carried out using nuclear emulsion and helium bubble chamber experiments beginning in the 1960s (see complete reference list in Ref. \cite{ada2018}). 

The first dedicated theoretical investigation of the three-body mesonic continuum was carried out by Motoba et al. \cite{mot1991, mot1992}. They developed a theoretical framework for calculating continuum pion spectra in the weak decays of light hypernuclei, including the hypertriton. Their work treated the continuum final states using the Kapur–Peierls formalism. This allowed for the calculation of pion energy spectra and partial decay rates, and  to investigate the influence of final-state interactions. 

A major development came several years later with the work of Kamada et al. \cite{kam1998}. They presented the first fully microscopic calculation of the hypertriton mesonic decay based on realistic three-body dynamics. In contrast to the approximate continuum treatment employed by Refs. \cite{mot1991, mot1992}, Kamada et al. solved the Faddeev equations. They incorporated realistic nucleon-nucleon and hyperon–nucleon interactions, including $\Lambda - \Sigma$ coupling, and accounted for final-state interactions among the outgoing nucleons. They obtained realistic initial-state hypertriton wave functions and rigorously treated the three-nucleon continuum in the final state. This work provided detailed predictions for partial decay rates, pion momentum spectra, branching ratios, and other differential observables. 

An important aspect of Ref. \cite{kam1998} is that it uses the Dalitz plot as a tool for visualizing the phase space of the $p+d+\pi^{-}$
breakup channel of hypertriton mesonic decay. Dalitz plots were first
applied in particle physics for the study of $K$-meson three-body decay \cite{dal1953, dal1954}. For a three-body decay $A \to 1+2+3$, the Dalitz plot visualizes the full final-state kinematics as a scatter of events in the two-dimensional plane of invariant-mass-squared combinations. Any two of the three pairwise invariant masses may be used, since the third is fixed by 4-momentum conservation. The pairwise invariant-mass-squared combinations conventionally used as Dalitz-plot axes are defined as

\begin{equation}
m^{2}_{12} \equiv (P_1 + P_2)^2
= (E_1+E_2)^2 - |\vec p_1 + \vec p_2|^2c^2
= m_1^2c^4 + m_2^2c^4 + 2E_1E_2 - 2\vec p_1\cdot\vec p_2\,c^2,
\label{eq:m12}
\end{equation}

\begin{equation}
m^{2}_{23} \equiv (P_2 + P_3)^2
= (E_2+E_3)^2 - |\vec p_2 + \vec p_3|^2c^2
= m_2^2c^4 + m_3^2c^4 + 2E_2E_3 - 2\vec p_2\cdot\vec p_3\,c^2,
\label{eq:m23}
\end{equation}

where $P_i=(E_i, \mathbf{p}_i)$ and $m_ic^2$ are the four-momentum and invariant mass of particle $i$. Dalitz plots provide a useful tool for investigating dynamical effects through deviations from a uniform phase-space distribution. However, individual momenta are the appropriate quantities for connecting to physical or experimental observations. They are also valuable for revealing structure that may not be immediately obvious from invariant-mass variables e.g. momentum correlation between pairs of decay products, opening angles, etc. In Ref. \cite{meo2026}, we developed a root-finding method that directly solves the three-body decay problem and determines individual momenta. In this method, the individual momenta are parametrised through a universal momentum variable, and a root-finding method is applied to solve the energy conservation equation in terms of this universal momentum variable. For a large number of events generated through a Monte Carlo sampling of momentum scaling parameters and the spherical angular space, this energy conservation equation is solved for each event, and the individual momenta ($p_1$, $p_2$ and $p_3$) are reconstructed.

In this paper, we present a comprehensive Raubold--Lynch phase-space construction \cite{jam1968, nav2019} for the three-body mesonic decay of the hypertriton in both the charged and neutral  pion channels. The algorithm factorises the Lorentz-invariant three-body phase space into two successive exact two-body decays through an intermediate off-shell  nucleon--deuteron subsystem. Complete final-state four-momenta are generated event by event while enforcing 4-momentum conservation. A key advantage of this approach is the separation of kinematics from dynamics: the generated phase-space samples can be directly weighted event by event through any weak-decay matrix element, without regenerating the kinematics.

\subsection{Three-body decay generated from Lorentz-invariant phase space}

\subsubsection{Sequential two-body decay construction}

In the Raubold-Lynch method, the three-body mesonic decay is generated using a recursive factorisation of the Lorentz-invariant three-body phase space \cite{jam1968}. Instead of treating the three final-state particles simultaneously, the decay is represented as two successive two-body decays through an intermediate virtual system,

\begin{align}
{}^3_\Lambda\mathrm H
\rightarrow
\pi^-+X,
\qquad
X
\rightarrow
p+d,
\end{align}

where $X$ is an off-shell particle representing the proton--deuteron subsystem. This intermediate particle does not correspond to a physical state. Rather, it is introduced solely to factorise the three-body phase space into two analytically solvable two-body decays. The four-momentum of this intermediate state is the sum of the proton and deuteron four-momenta,

\begin{equation}
P_X=P_p+P_d,
\end{equation}

where $P_p=(E_p,\mathbf{p}_p)$ and $P_d=(E_d,\mathbf{p}_d)$. The invariant mass of the intermediate system is therefore

\begin{equation}
m_{pd}
=
\sqrt{P_X^2}
=
\sqrt{(E_p+E_d)^2-
\left|\mathbf{p}_p+\mathbf{p}_d\right|^2}.
\end{equation}

Unlike a two-body decay, where the daughter masses uniquely determine the final-state kinematics, a three-body decay possesses one additional independent kinematic degree of freedom. Consequently, the proton and deuteron do not always share the available decay energy in the same way, and the invariant mass of their combined system varies continuously from one decay event to another. Physically, different values of $m_{pd}$ correspond to different partitions of the available decay energy among the proton, deuteron and pion while satisfying 4-momentum conservation.

The allowed range of $m_{pd}$ is determined entirely by relativistic kinematics. The minimum value occurs when the proton and deuteron are produced with zero relative momentum in their common centre-of-mass frame. In this threshold configuration,

\begin{equation}
m_{pd}^{\mathrm{min}}
=
m_p+M_d.
\end{equation}

The maximum value is attained when the pion has its minimum possible
energy in the hypertriton rest frame, allowing the proton--deuteron subsystem to attain its largest possible invariant mass,

\begin{equation}
m_{pd}^{\mathrm{max}}
=
M_0-m_{\pi^-},
\end{equation}

where $M_0$ denotes the hypertriton mass. The parent mass $M_0$ is fixed directly by the
three-body $Q$-value,
\begin{equation}
  M_0 \;=\; M_d + m_p + m_{\pi^-} + Q^{3}_{\pi^-},
  \label{eq:M0_definition}
\end{equation}
with $Q^{3}_{\pi^-} = Q_{\pi^-} - B_{\Lambda}= 37.32$~MeV, where $Q_{\pi^-} = 37.841$ MeV is the Q-value of the free lambda mesonic decay and $B_{\Lambda}=0.523$ MeV is the lambda separation energy from the MAMI A1 experiment \cite{kin2026}. As derived in Ref. \cite{meo2026}, the Q-value for the two-body decay is given by $Q^{2}_{\pi^-} = Q_{\pi^-} - B_{\Lambda} + S_p = 42.81$ MeV, where $S_p=  5.4934 $ MeV is the proton separation energy of the 3He. Therefore, the energy that is available for sharing between the decay products is smaller in three-body decay (37.32 MeV) than for two-body decay (42.81 MeV).

The kinematically allowed range for $m_{pd} $ is

\begin{equation}
m_p+M_d
\le
m_{pd}
\le
M_0-m_{\pi^-}.
\end{equation}

In the Raubold--Lynch algorithm used here, the invariant mass $m_{pd}$ is sampled from its kinematically allowed range and each event is assigned a phase-space weight

\begin{equation}
  w \;\propto\; p^{*}(M_0, m_{\pi^-}, m_{pd})\,
                p^{**}(m_{pd}, m_p, M_d),
  \label{eq:genbod_weight}
\end{equation}

This is the product of the two-body breakup momenta from the
$^{3}_{\Lambda}\mathrm{H} \to \pi^- + X$ and $X \to p + d$ stages. Once a value of $m_{pd}$ has been generated, the first decay,

\begin{align}
{}^3_\Lambda\mathrm H
\rightarrow
\pi^-+X(m_{pd}),
\end{align}

is completely determined by two-body kinematics. The resulting four-momentum of the intermediate system $X$ then specifies the kinematics of the second decay,

\begin{align}
X(m_{pd})
\rightarrow
p+d,
\end{align}
which is likewise solved analytically using two-body kinematics. Repeating this procedure, the algorithm generates kinematically allowed configurations using a distribution in $m_{pd} $ and isotropic angular variables. The phase-space weight is then used to obtain an ensemble distributed according to the Lorentz-invariant three-body phase-space measure. The invariant mass $m_{pd}$ therefore serves as the single independent invariant-mass variable governing the sequential decomposition of the decay.

\subsubsection{Two-body kinematics and angular generation}

Once a value of $m_{pd}$ has been generated, each stage of the sequential decay becomes an ordinary two-body process with closed-form relativistic kinematics. The first stage,

\begin{align}
{}^3_{\Lambda}\mathrm H
\rightarrow
\pi^-+X(m_{pd}),
\end{align}

takes place in the hypertriton rest frame. The pion and the intermediate system $X$ recoil back-to-back with equal momentum magnitude

\begin{align}
\label{eq:breakup_one}
p_{\pi}
=
p_X
=
\frac{1}{2M_0}
\sqrt{\left[M_0^2-(m_{\pi^-}+m_{pd})^2\right]
      \left[M_0^2-(m_{\pi^-}-m_{pd})^2\right]},
\end{align}

where $M_0$ is the hypertriton mass. Their corresponding energies are

\begin{align}
E_{\pi}
&=
\sqrt{m_{\pi^-}^{\,2}+p_{\pi}^{\,2}},
\\
E_X
&=
\sqrt{m_{pd}^{\,2}+p_X^{\,2}}.
\end{align}

At this stage, only the magnitude of the momentum is known. The direction of emission remains undetermined. For an unpolarized hypertriton and in the absence of dynamical angular correlations, the decay direction is isotropically distributed. The isotropic orientation is generated as a composition of two elementary axis rotations applied to an axis-aligned configuration. The pion and the intermediate system are first placed along a common reference axis,

\begin{equation}
\mathbf{p}_\pi^{(0)} = (0,\,p_\pi,\,0), \qquad \mathbf{p}_X^{(0)} = (0,\,-p_\pi,\,0),
\end{equation}

with $p_\pi$ as already computed. An isotropic direction is then obtained by two sequential random rotations \cite{kyc2019}:

\begin{enumerate}
    \item Rotation about the $z$-axis, by angle $\beta_z^{(1)}$, where
\begin{equation}
\cos\beta_z^{(1)} \in [-1,1] \ \text{(uniform)}, \qquad
\sin\beta_z^{(1)} = \sqrt{1-\cos^2\beta_z^{(1)}},
\end{equation}
\begin{equation}
p_x' = \cos\beta_z^{(1)}\,p_x - \sin\beta_z^{(1)}\,p_y, \qquad
p_y' = \sin\beta_z^{(1)}\,p_x + \cos\beta_z^{(1)}\,p_y, \qquad
p_z' = p_z.
\end{equation}

\item Rotation about the $y$-axis, by angle $\beta_y^{(1)} \in [0,2\pi)$ (uniform),
\begin{equation}
p_x'' = \cos\beta_y^{(1)}\,p_x' - \sin\beta_y^{(1)}\,p_z', \qquad
p_z'' = \sin\beta_y^{(1)}\,p_x' + \cos\beta_y^{(1)}\,p_z', \qquad
p_y'' = p_y'.
\end{equation}

\end{enumerate}

Applying these rotations to $\mathbf{p}_\pi^{(0)}$ gives the oriented pion momentum,
\begin{equation}
\mathbf{p}_\pi = (p_x'', p_y'', p_z''), \qquad P_\pi = (E_\pi, \mathbf{p}_\pi),
\end{equation}
while momentum conservation requires $\mathbf{p}_X = -\mathbf{p}_\pi$, so that
\begin{equation}
P_X = (E_X, -\mathbf{p}_\pi).
\end{equation}

The second stage,

\begin{align}
X(m_{pd})
\rightarrow
p+d,
\end{align}

is performed in the rest frame of the intermediate system. The proton and deuteron again recoil back-to-back with common momentum magnitude

\begin{align}
\label{eq:breakup_two}
p_p
=
p_d
=
\frac{1}{2m_{pd}}
\sqrt{\left[m_{pd}^{\,2}-(m_p+M_d)^2\right]
      \left[m_{pd}^{\,2}-(m_p-M_d)^2\right]}.
\end{align}

Their corresponding energies are

\begin{align}
E_p
&=
\sqrt{m_p^{\,2}+p_p^{\,2}},
\\
E_d
&=
\sqrt{M_d^{\,2}+p_d^{\,2}}.
\end{align}

As in the first decay, only the momentum magnitude of the outgoing particles is fixed. Once a value of $m_{pd}$ has been generated, the second decay stage,
\begin{equation}
X(m_{pd}) \to p + d,
\end{equation}
is likewise generated using the two-rotation (z-then-y) isotropic construction, applied this time
to the proton--deuteron subsystem in the rest frame of $X$. The proton and deuteron are first
placed along a common reference axis,

\begin{equation}
\mathbf{p}_p^{(0)} = (0,\,p_p,\,0), \qquad \mathbf{p}_d^{(0)} = (0,\,-p_p,\,0),
\end{equation}

with $p_p$ as already computed. Two independent random angles, $\beta_z^{(2)}$ and $\beta_y^{(2)}$,
are drawn for this stage:

\begin{enumerate}
    \item Rotation about the $z$-axis, by angle $\beta_z^{(2)}$, where
\begin{equation}
\cos\beta_z^{(2)} \in [-1,1] \ \text{(uniform)}, \qquad
\sin\beta_z^{(2)} = \sqrt{1-\cos^2\beta_z^{(2)}},
\end{equation}
\begin{equation}
p_x' = \cos\beta_z^{(2)}\,p_x - \sin\beta_z^{(2)}\,p_y, \qquad
p_y' = \sin\beta_z^{(2)}\,p_x + \cos\beta_z^{(2)}\,p_y, \qquad
p_z' = p_z.
\end{equation}

\item Rotation about the $y$-axis, by angle $\beta_y^{(2)} \in [0,2\pi)$ (uniform),
\begin{equation}
p_x'' = \cos\beta_y^{(2)}\,p_x' - \sin\beta_y^{(2)}\,p_z', \qquad
p_z'' = \sin\beta_y^{(2)}\,p_x' + \cos\beta_y^{(2)}\,p_z', \qquad
p_y'' = p_y'.
\end{equation}

\end{enumerate}

Applying these rotations to $\mathbf{p}_p^{(0)}$ gives the oriented proton momentum,
\begin{equation}
\mathbf{p}_p = (p_x'', p_y'', p_z''), \qquad P_p = (E_p, \mathbf{p}_p),
\end{equation}
while the deuteron recoils in the opposite direction,
\begin{equation}
P_d = (E_d, -\mathbf{p}_p).
\end{equation}

Thus, for each stage of the Raubold--Lynch construction, the breakup momentum obtained from
relativistic two-body kinematics determines the magnitude of the daughter momenta, whereas the randomly sampled rotation angles $(\beta_z,\beta_y)$ (a $z$-rotation with $\cos\beta_z$ uniform in $[-1,1]$ followed by a $y$-rotation with $\beta_y$ uniform in $[0,2\pi)$) determine their orientation in three-dimensional space. This two-rotation composition distributes the resulting direction uniformly over the unit sphere. Together, the breakup momentum and the sampled rotation angles uniquely determine the complete daughter four-momenta before the proton and deuteron are Lorentz boosted into the hypertriton rest frame.

\subsubsection{Lorentz transformation}

At the end of the previous step, the pion four-momentum $P_{\pi}$ has already been constructed in the hypertriton rest frame, whereas the proton and deuteron four-momenta, $P_p$ and $P_d$, are still expressed in the rest frame of the intermediate system $X$. To obtain the complete three-body final state in a common reference frame, the proton and deuteron four-momenta must therefore be Lorentz boosted into the hypertriton rest frame.

The boost is completely determined by the motion of the intermediate system produced in the first decay,
\begin{align}
{}^3_{\Lambda}\mathrm H
\rightarrow
\pi^-+X,
\end{align}
whose four-momentum was obtained in the preceding subsection. The corresponding boost velocity is

\begin{equation}
\boldsymbol{\beta}_X
=
\frac{\mathbf p_X}{E_X},
\qquad
\gamma_X
=
\frac{1}{\sqrt{1-|\boldsymbol{\beta}_X|^2}}
=
\frac{E_X}{m_{pd}},
\end{equation}

where $E_X$ and $\mathbf p_X$ are the energy and three-momentum of the intermediate system in the hypertriton rest frame, and $m_{pd}$ is the sampled invariant mass of the proton--deuteron subsystem.

For either daughter particle, $i\in\{p,d\}$, with four-momentum

\[
P_i=(E_i,\mathbf p_i)
\]

defined in the rest frame of $X$, the Lorentz transformation to the hypertriton rest frame gives \cite{jac1998}

\begin{align}
E_i' &=\gamma_X\left(E_i + \boldsymbol{\beta}_X\cdot\mathbf p_i \right), \\
\mathbf p_i' &= \mathbf p_i + \left[ (\gamma_X-1) \frac{\mathbf p_i\cdot\boldsymbol{\beta}_X} {|\boldsymbol{\beta}_X|^2}
+
\gamma_XE_i
\right]
\boldsymbol{\beta}_X.
\end{align}

Applying these equations to both the proton and the deuteron transforms their four-momenta from the rest frame of the intermediate system into the hypertriton rest frame while preserving the invariant mass of the $(p,d)$ subsystem. Since Lorentz transformations preserve four-momentum conservation, the boosted four-vectors satisfy

\begin{equation}
P_p'+P_d'
=
P_X,
\end{equation}

where $P_X$ is the four-momentum of the intermediate system obtained from the first decay stage. Together with the pion four-momentum, which already resides in the hypertriton rest frame and therefore requires no further transformation, the complete three-body final state is therefore specified by

\begin{equation}
P_{{}^3_{\Lambda}\mathrm H}
=
P_{\pi}
+
P_p'
+
P_d',
\end{equation}

which satisfies 4-momentum conservation. The four-vectors $P_{\pi}$, $P_p'$, and $P_d'$ therefore constitute the final kinematic configuration of a single three-body decay event and form the basis for the calculation of momentum spectra, angular distributions, and Dalitz plots presented in the results section.

\subsubsection{Extraction of Momentum Observables}

Following the Lorentz transformation, the four-momenta of all three decay products, $P_{\pi}, P_p'$ and $ P_d'$, are now expressed in the hypertriton rest frame. These four-vectors completely specify the kinematics of a single decay event and provide the basis for calculating the physical observables of interest.

For each decay product, $i\in\{\pi^-,p,d\}$, the magnitude of the three-momentum is obtained from its Cartesian components,

\begin{equation}
|\mathbf{p}_i|
=
\sqrt{p_{i,x}^{\,2}+p_{i,y}^{\,2}+p_{i,z}^{\,2}},
\end{equation}

while the corresponding kinetic energy is calculated as

\begin{equation}
T_i
=
E_i-m_i,
\end{equation}

where $E_i$ and $m_i$ denote the particle energy and rest mass, respectively.

Repeating the Raubold--Lynch procedure over a large ensemble of
Monte Carlo events produces a sample distributed according to the
Lorentz-invariant three-body phase-space measure.

The event-by-event values of $|\mathbf p_i|$ and $T_i$ are subsequently accumulated to construct the momentum and kinetic-energy spectra of the pion, proton, and deuteron. The generated four-momenta are also used to calculate angular distributions and Dalitz plots, providing a complete kinematic description of the decay. Because the Raubold--Lynch algorithm generates only the available phase space, these events can subsequently be weighted with realistic weak-decay matrix elements to incorporate the underlying decay dynamics while preserving the exact relativistic kinematics established by the phase-space construction.

\section{From phase space to decay rate: the role of this calculation}

The Raubold--Lynch construction developed in the foregoing section solves a purely kinematic problem. For a given three-body $Q$-value $Q_{\pi^-}^3$ and daughter masses, it generates the full set of momentum configurations $(\mathbf{p}_{\pi^-},\mathbf{p}_p,\mathbf{p}_d)$ allowed by energy--momentum conservation. The momenta are distributed according to the exact Lorentz-invariant three-body phase-space element \cite{pat2016}

\begin{equation}
d\Phi_3(P;p_\pi,p_p,p_d)=(2\pi)^4\delta^{(4)}(P-p_\pi-p_p-p_d)\prod_{i=\pi,p,d}\frac{d^3p_i}{(2\pi)^3 2E_i}.
\label{eq:phi3}
\end{equation}

This phase space factorises through the intermediate $(p,d)$ subsystem as

\begin{equation}
d\Phi_3=\frac{1}{2\pi}\,dm_{pd}^2\,d\Phi_2({}^3_\Lambda\mathrm{H}\to X+\pi^-)\,d\Phi_2(X\to p+d),
\label{eq:phi3factorized}
\end{equation}

which is exactly the sequential two-body decomposition implemented by the algorithm. Each factor on the right is an ordinary two-body Lorentz-invariant phase-space element. For a parent of mass $M$ decaying into two particles with common breakup momentum $p^*$, this two-body element takes the closed form \cite{pat2016}

\begin{equation}
d\Phi_2(P;p_1,p_2)=(2\pi)^4\delta^{(4)}(P-p_1-p_2)\prod_{i=1,2}\frac{d^3p_i}{(2\pi)^3 2E_i}=\frac{p^*}{16\pi^2 M}\,d\Omega^*,
\label{eq:phi2}
\end{equation}

with $d\Omega^*$ the solid angle of either daughter in the two-body rest frame. Applied to the two stages of Eq.~\eqref{eq:phi3factorized}, this gives explicitly

\begin{equation}
d\Phi_2({}^3_\Lambda\mathrm{H}\to X+\pi^-)=\frac{p_1^*(m_{pd})}{16\pi^2 M_0}\,d\Omega_\pi,
\qquad
d\Phi_2(X\to p+d)=\frac{p_2^*(m_{pd})}{16\pi^2 m_{pd}}\,d\Omega_p,
\label{eq:phi2explicit}
\end{equation}

where $p_1^*(m_{pd})=p_\pi$ is the breakup momentum of Eq.~\ref{eq:breakup_one} evaluated in the ${}^3_\Lambda\mathrm{H}$ rest frame, and $p_2^*(m_{pd})=p_p$ is the breakup momentum of Eq.~\ref{eq:breakup_two} evaluated in the rest frame of $X$. Since the decay is isotropic at each stage, integrating each factor over its full solid angle gives the angle-integrated two-body phase space $\Phi_2=p^*/(4\pi M)$, so that

\begin{equation}
\frac{d\Phi_3}{dm_{pd}^2}=\frac{1}{2\pi}\left[\frac{p_1^*(m_{pd})}{4\pi M_0}\right]\left[\frac{p_2^*(m_{pd})}{4\pi m_{pd}}\right]=\frac{p_1^*(m_{pd})\,p_2^*(m_{pd})}{32\pi^3 M_0\, m_{pd}}.
\label{eq:phi3over_mpdsq}
\end{equation}

Converting from $m_{pd}^2$ to $m_{pd}$ via $dm_{pd}^2=2m_{pd}\,dm_{pd}$ then reproduces the differential phase-space density,

\begin{equation}
\frac{d\Phi_3}{dm_{pd}}=\frac{p_1^*(m_{pd})\,p_2^*(m_{pd})}{16\pi^3 M_0},
\label{eq:dphidmpd}
\end{equation} 

The observable of ultimate interest is not $d\Phi_3$ itself, but the decay rate. The physical decay rate is obtained by weighting the Lorentz-invariant phase-space measure by the squared transition matrix element. In the standard particle-physics convention ($\hbar = c = 1$), the differential three-body decay rate of the hypertriton is

\begin{equation}
d\Gamma = \frac{1}{2M}\,\overline{|\mathcal{M}(\mathbf{p}_{\pi^-},\mathbf{p}_p,\mathbf{p}_d)|^2}\; d\Phi_3(P;p_{\pi},p_p,p_d),
\label{eq:dgamma}
\end{equation}

where $M$ is the hypertriton mass, $\mathcal{M}$ is the weak-decay transition amplitude connecting the initial ${}^3_\Lambda\mathrm{H}$ bound state to the $\pi^- p d$ continuum, and the bar denotes the appropriate spin average/sum. The distributions shown in the results section are Monte Carlo realizations of the Lorentz-invariant phase-space measure $d\Phi_3$ without dynamical matrix-element weighting.

The Raubold--Lynch method yields, for each event, the full on-shell four-momenta of all three decay products, together with a phase-space weight $w\propto p_1^*(m_{pd})\,p_2^*(m_{pd})$. The accept/reject step converts this weighted sample into a bona fide unweighted phase-space sample. Crucially, the full four-vector sample is retained and written to disk, together with the sampled $m_{pd}$ value and weight for every event.

\subsection{Accept/reject procedure}

The phase-space weight quoted earlier, $w\propto p_1^*(m_{pd})\,p_2^*(m_{pd})$, is made explicit for the accept/reject step by normalizing to its maximum over the raw sample of $N_{\text{raw}}=10^5$ draws,

\begin{equation}
w_i
= \frac{p_1^*(m_{pd,i})\,p_2^*(m_{pd,i})}
       {\max\limits_{j=1,\dots,N_{\text{raw}}}\big[p_1^*(m_{pd,j})\,p_2^*(m_{pd,j})\big]}
\in[0,1],
\label{eq:weight_exact}
\end{equation}

Each raw event $i$ is kept if $r_i<w_i$, with $r_i\sim\mathcal U(0,1)$, and rejected otherwise. In this study we used $10^5$ raw weighted events. This accept/reject procedure reduces these weighted draws to the $37923$ unweighted events that are used in the results section. The normalization in the denominator of Eq.~\eqref{eq:weight_exact} is a numerical device tied to the finite raw sample. It fixes the acceptance efficiency of the
Monte Carlo run.

\subsection{Monte Carlo integration over phase space.}

The Raubold–Lynch construction generates kinematically allowed three-body configurations from a proposal distribution in the invariant mass and angular variables. The phase-space weight is then used in an accept/reject procedure to obtain an unweighted sample distributed according to $d\Phi_3$. 

After the accept/reject procedure, the accepted events form an unweighted Monte Carlo sample distributed according to the normalized Lorentz-invariant three-body phase-space measure,

\begin{equation}
dP(\Phi_3)
=
\frac{d\Phi_3}{\Phi_3^{\mathrm{tot}}},
\qquad
\Phi_3^{\mathrm{tot}}=\int d\Phi_3.
\end{equation}

Here, $dP$ is the probability measure on the full three-body phase space.  When expressed with respect to a particular kinematic coordinate $x$, the corresponding probability density is given by

\begin{equation}
f(x)
\equiv
\frac{dP}{dx}
=
\frac{1}{\Phi_3^{\mathrm{tot}}}
\frac{d\Phi_3}{dx},
\end{equation}

with the normalization

\begin{equation}
\int f(x)\,dx=1.
\end{equation}

Once the decay matrix element $\mathcal{M}(\Phi_3)$
is specified, the total three-body decay width is

\begin{align}
\Gamma_3^{\pi^-}
=
\frac{1}{2M_0}
\int
\overline{|\mathcal{M}|^2}\,d\Phi_3 = \frac{\Phi_3^{\text{tot}}}{2M_0} \int
\overline{|\mathcal{M}|^2}\,dP
\label{eq:Gamma_int}
\end{align}

where the bar denotes the appropriate average over the initial spin
and sum over the final spins. The integral is the expectation value of $|\mathcal{M}|^2$ i.e.

\begin{equation}
\Gamma_3^{\pi^-}
=
\frac{\Phi_3^{\text{tot}}}{2M_0}
\left\langle
\overline{|\mathcal{M}|^2}
\right\rangle_{\Phi_3}.
\label{eq:Gamma_average}
\end{equation}

For an unweighted sample of $N$ independent phase-space events, and applying the law of large numbers on the expectation value, the decay rate is given by \cite{byc1973, jam1968, jam1980}
\begin{equation}
\Gamma_3^{\pi^-}
\simeq
\frac{\Phi_3^{\text{tot}}}{2M_0}
\frac{1}{N}
\sum_{i=1}^{N}
\overline{|\mathcal{M}_i|^2}.
\label{eq:reweight}
\end{equation}
Here $\overline{|\mathcal{M}_i|^2}$ is the spin-averaged squared matrix
element evaluated using the four-momenta of event $i$.

The total three-body phase-space volume for the
$\pi^-pd$ final state is
\begin{equation}
\Phi_3^{\text{tot}}
=
\int d\Phi_3
=
\int_{m_p+M_d}^{\,M_0-m_{\pi^-}}
\frac{d\Phi_3}{dm_{pd}}\,dm_{pd}
=
\int_{m_p+M_d}^{\,M_0-m_{\pi^-}}
\frac{p_1^*(m_{pd})\,p_2^*(m_{pd})}
{16\pi^3 M_0}\,dm_{pd},
\label{eq:Phi3tot}
\end{equation}
where $p_1^*(m_{pd})$ is the momentum of the pion in the
hypertriton rest frame and $p_2^*(m_{pd})$ is the momentum of the proton
(or deuteron) in the $pd$ rest frame.

Equation~\eqref{eq:reweight} is therefore an ordinary Monte Carlo estimator of the phase-space integral in Eq.~\eqref{eq:Gamma_int}. Because the events are unweighted and distributed according to the normalized Lorentz-invariant phase-space measure $d\Phi_3/\Phi_3^{\text{tot}}$, no additional event-by-event phase-space weight is required. The estimator is unbiased in the limit of independent sampling and converges to the exact result as $N\rightarrow\infty$, with its statistical uncertainty decreasing as $1/\sqrt{N}$.

This strategy provides a foundation for extending the present work to include dynamical effects. A dynamical input can be developed and validated independently of the phase-space generator. Refs. \cite{kam1998, hil2020} are examples of matrix elements that may be implemented with the generated phase space of this paper. This strategy also allows for a reweighting of the sample event-by-event, as widely applied in particle physics \cite{cot2004, ber2026}.

\section{Results and discussion}

\subsection{Negative pion mesonic decay}

\begin{figure}[htbp]
    \centering
    \includegraphics[width=0.6\textwidth]{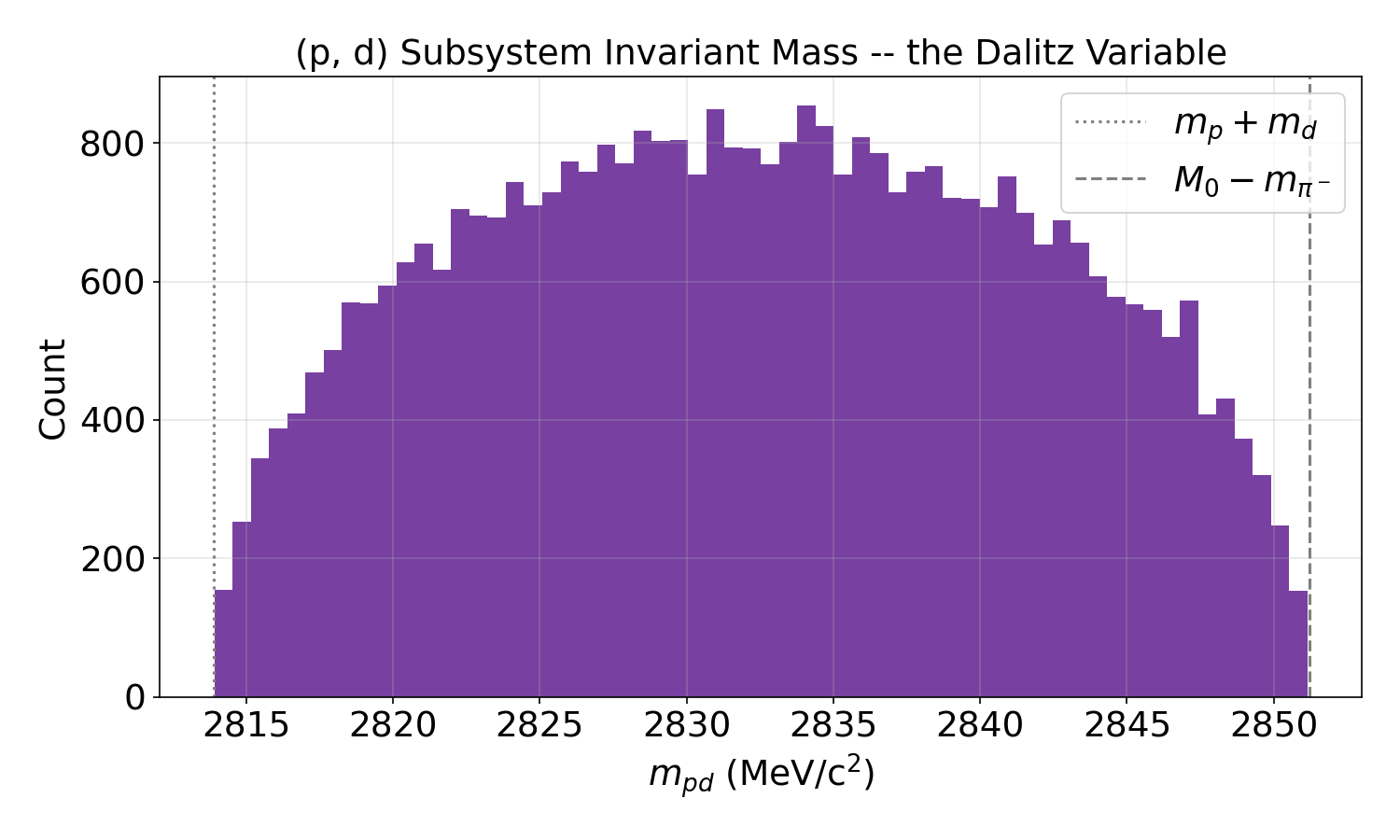}
    \caption{(p,d) subsystem invariant mass ($m_{pd}$) distribution. The minimum is 2813.89 and the maximum is 2851.21 MeV/$c^2$, with mean = 2832.44 MeV/$c^2$, standard deviation = 9.34 MeV/$c^2$.}
    \label{fig:m_pd_invariant_mass}
\end{figure}

Figure~\ref{fig:m_pd_invariant_mass} shows the distribution of the proton--deuteron invariant mass, $m_{pd}$, generated by the Raubold--Lynch phase-space algorithm. The quantity plotted in Fig.~\ref{fig:m_pd_invariant_mass} is the invariant mass, \(m_{pd}\), which is restricted by energy conservation to

\[
m_p+m_d=2813.9~\mathrm{MeV}/c^2
\le
m_{pd}
\le
M_0-m_{\pi^-}=2851.2~\mathrm{MeV}/c^2,
\]

The vertical axis gives the raw number of Monte Carlo events (Count) contained in each of the 60 invariant-mass bins. As shown earlier, the sequential decay

\[
{}^3_\Lambda\mathrm{H}
\rightarrow
X+\pi^-,
\qquad
X\rightarrow p+d,
\]

factorises the three-body Lorentz-invariant phase-space element $d\Phi_3$ into a product of two-body factors, yielding the differential phase-space density of

\begin{equation}
\frac{d\Phi_3}{dm_{pd}}
=
\frac{
p_1^{*}(m_{pd})\,
p_2^{*}(m_{pd})
}
{16\pi^3M_0},
\label{eq:dphi3-recall}
\end{equation}
where \(p_1^{*}\) is the pion momentum in the hypertriton rest frame and
\(p_2^{*}\) is the proton momentum in the rest frame of the proton--deuteron subsystem. Numerically, Eq.~\eqref{eq:Phi3tot} evaluates to
$\Phi_3^{\text{tot}} = 2.194747\times10^{-1}\ \text{MeV}^2$, stored
alongside the event sample as the normalization needed for
Eq.~\eqref{eq:reweight}.

Eq.~(\ref{eq:dphi3-recall}) establishes the connection between the theoretical phase-space density and the histogram in Fig.~\ref{fig:m_pd_invariant_mass}. The number of events recorded in each bin is determined by the value of \(d\Phi_3/dm_{pd}\) over that interval. Near the lower kinematic limit,
\(m_{pd}=m_p+m_d\),
the proton and deuteron have vanishing relative momentum so that
\(p_2^{*}\rightarrow0\). At the upper limit,
\(m_{pd}=M_0-m_{\pi^-}\),
the pion momentum approaches zero,
\(p_1^{*}\rightarrow0\). Consequently,
\[
\frac{d\Phi_3}{dm_{pd}}
\rightarrow
0
\]
at both endpoints, producing the low event counts observed at the edges of the histogram. Between these limits both breakup momenta are appreciable, the phase-space density is largest, and the histogram therefore exhibits its highest event counts.

The distribution in Fig.~\ref{fig:m_pd_invariant_mass} represents pure Lorentz-invariant phase space. Any physical decay dynamics, such as the hypertriton wave function or proton--deuteron final-state interactions, would enter through $|\overline{\mathcal{M}}(p_{\pi^-},p_p,p_d)|^2$, thereby enhancing or suppressing specific regions of the invariant-mass spectrum relative to the flat baseline shown here.

\subsubsection{Normalised probability density and histogram of individual momenta} 

\begin{figure}[htbp]
    \centering
    \begin{subfigure}[b]{0.48\textwidth}
        \centering
        \includegraphics[width=\textwidth]{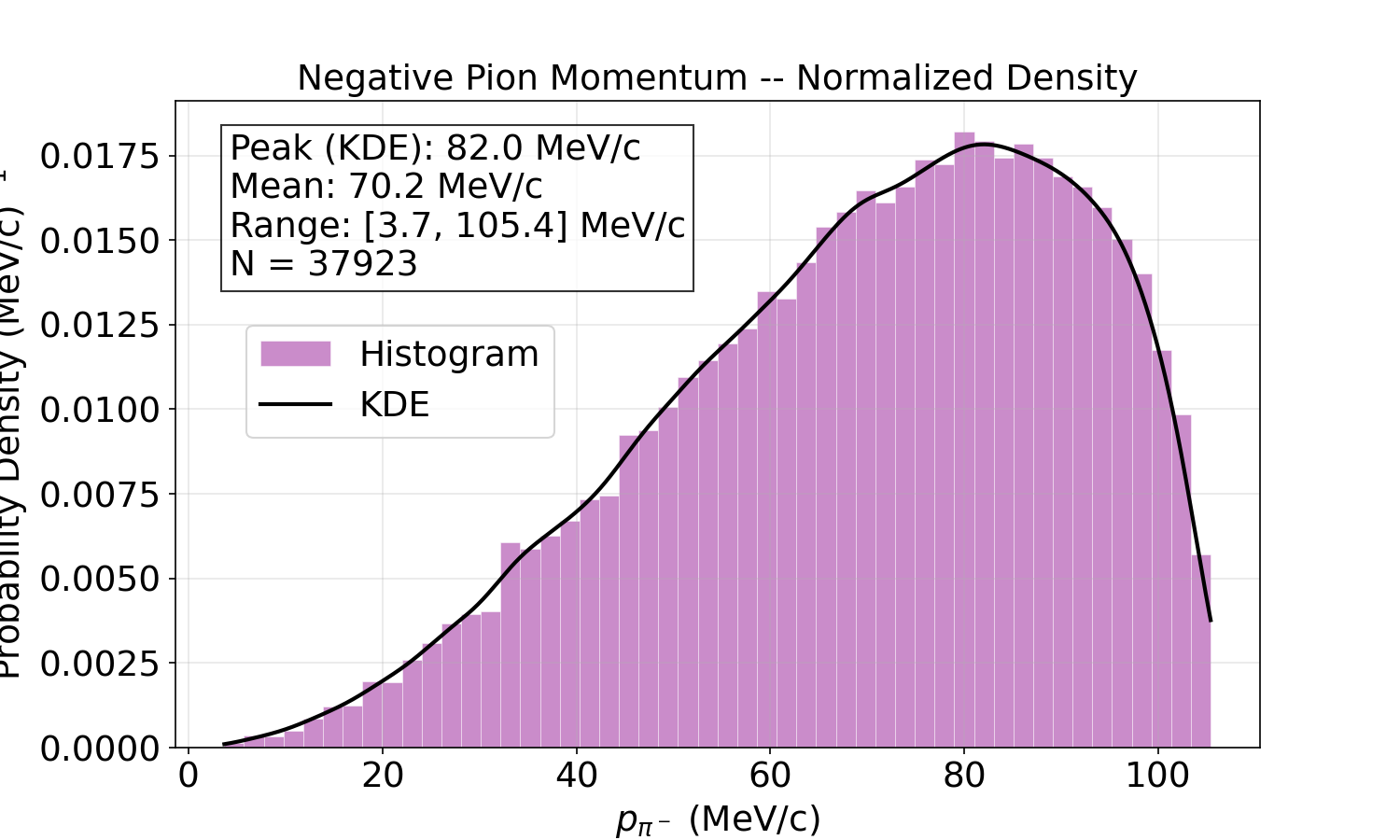}
        \caption{Pion momentum spectrum. Max = 105.39, mean = 70.18 MeV/c.}
        \label{fig:pion_spectrum}
    \end{subfigure}
    \hfill
    \begin{subfigure}[b]{0.48\textwidth}
        \centering
        \includegraphics[width=\textwidth]{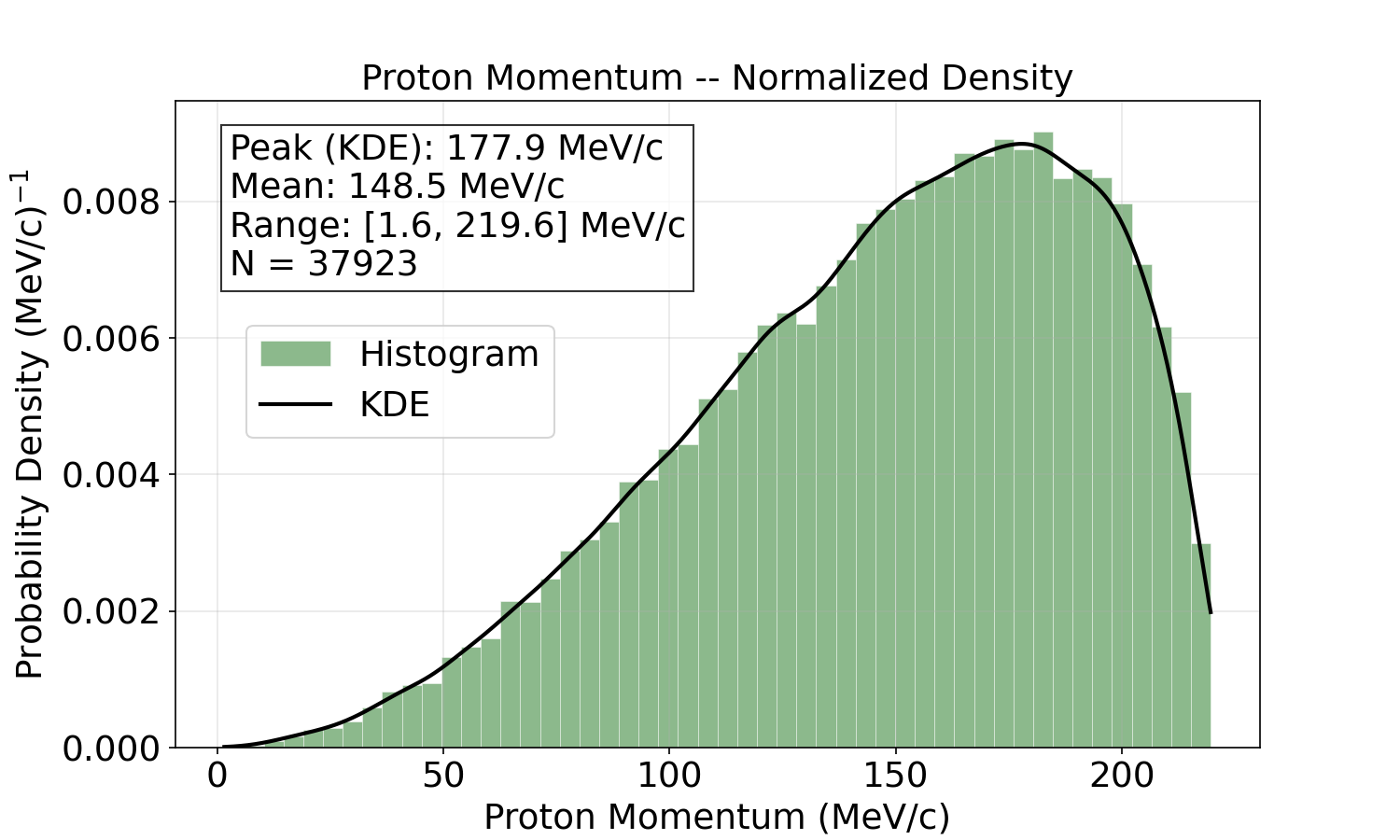}
        \caption{Proton momentum spectrum. Max = 219.57, mean = 148.53 MeV/c}
        \label{fig:proton_spectrum}
    \end{subfigure}

    \vspace{1em}

    \begin{subfigure}[b]{0.48\textwidth}
        \centering
        \includegraphics[width=\textwidth]{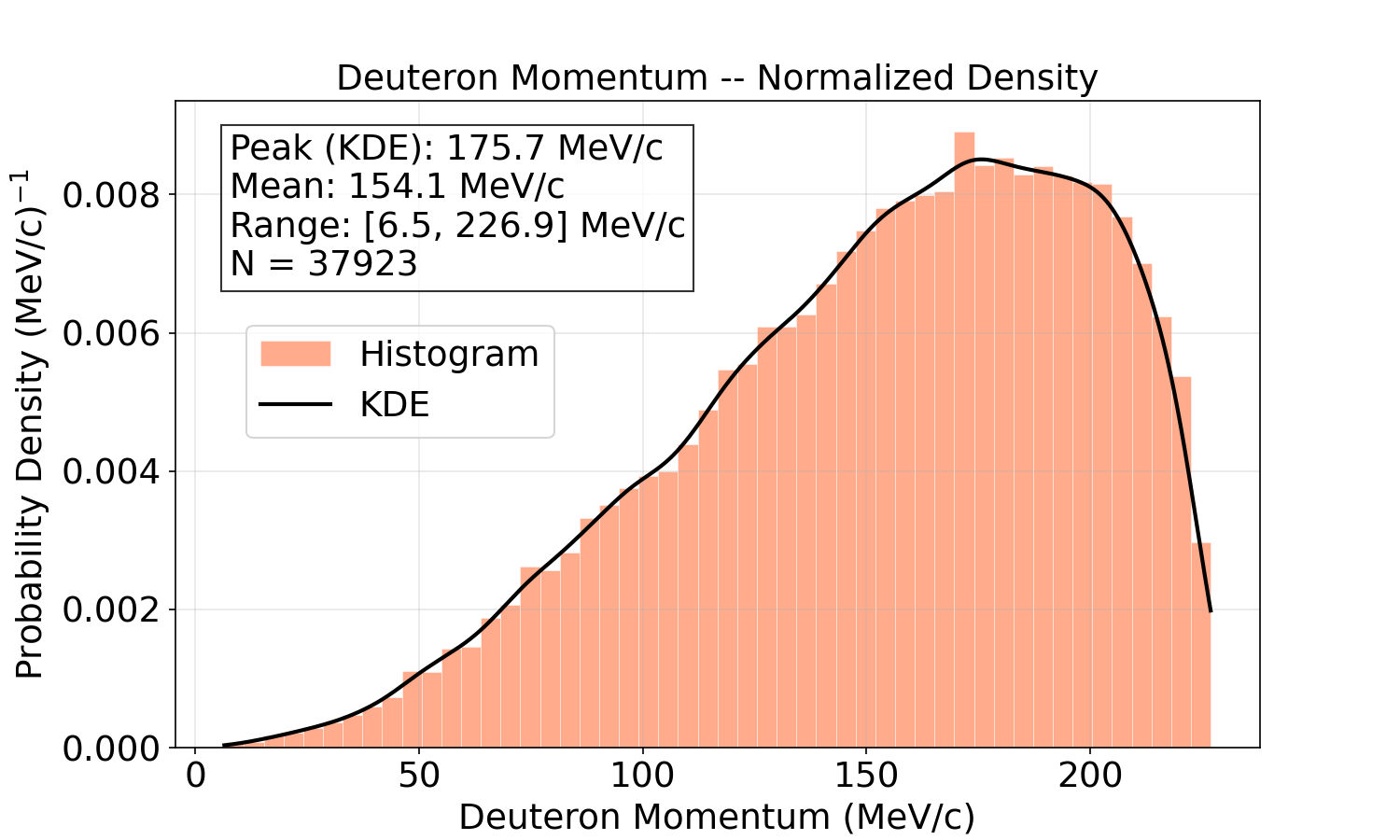}
        \caption{Deuteron momentum spectrum. Max = 226.92, mean = 154.11 MeV/c.}
        \label{fig:deuteron_spectrum}
    \end{subfigure}

    \caption{Normalised momentum spectra for the three decay products of
    $^3_\Lambda\text{H} \to p + d + \pi^-$, generated using the
    Raubold--Lynch construction.}
    \label{fig:momentum_spectra}
\end{figure}

Raw weighted events were converted into a physical, unweighted sample via an accept/reject procedure, yielding $37923$ kinematically valid events used in all subsequent analysis. Figure \ref{fig:momentum_spectra} shows a smooth kernel density estimate (KDE) overlaid on a histogram of each distribution. The three momentum spectra reveal several physically meaningful features of the decay. 

Each momentum distribution has a pronounced asymmetry, i.e., a long low-momentum tail and an abrupt high-momentum cut-off. All three spectra rise smoothly from near-zero momentum, reach a broad maximum, and fall sharply to a hard upper kinematic limit. The upper limits for each distribution are \(105.39\,\mathrm{MeV}/c\) for the pion, \( 219.57\,\mathrm{MeV}/c\) for the proton and \( 226.92\,\mathrm{MeV}/c\) for the deuteron. The upper limit for each particle corresponds to the configuration where that particle recoils against a two-body subsystem of the remaining two particles together. Each particle's minimum-momentum configuration ($p_i \to 0$ ) is the "two-body-like" limit where that particle drops out of the dynamics and the remaining pair shares all the kinetic energy between themselves. These limits are model-independent once \(Q_{\pi^-}^3  \) is chosen. Dynamical effects cannot change these kinematically allowed ranges of momenta. Rather, they redistribute events within these limits.

The pion exhibits the smallest mean momentum (with a peak of \(82\,\mathrm{MeV}/c\)) in the present phase-space distribution, while the proton and deuteron have substantially larger mean momenta. The proton and deuteron exhibit nearly identical spectra (peaks \( 177.9\,\mathrm{MeV}/c\) and \( 175.7\,\mathrm{MeV}/c\), respectively). The similarity of the proton and deuteron spectra reflects their participation in the recoiling $p-d$ subsystem, while their small differences arise from the unequal masses and the three-body kinematics.

The pion-momentum endpoint of the three-body continuum,
$p_{\pi^-}^{\max}=105.39$~MeV/c, lies $8.4$~MeV/c below the two-body
monochromatic momentum, $p_{\pi^-}=113.789$~MeV/c, that was recently measured by the MAMI A1 collaboration \cite{kin2026}. Given the relative momentum resolution of the A1 spectrometers, the three-body $d+p+\pi^-$ continuum can therefore be considered kinematically well-separated from the two-body peak. It is impossible for the continuum's endpoint tail to leak into the fitted two-body line. This conclusion remains unchanged even if dynamical effects are included in the study of the three-body continuum through a weak-decay matrix element. Dynamical effects only change the way events populate the phase space, not the kinematic limits of the phase space.

\subsubsection{Pairwise momentum correlation between decay products}

\begin{figure}[htbp]
    \centering
    \begin{subfigure}[b]{0.48\textwidth}
        \centering
        \includegraphics[width=\textwidth]{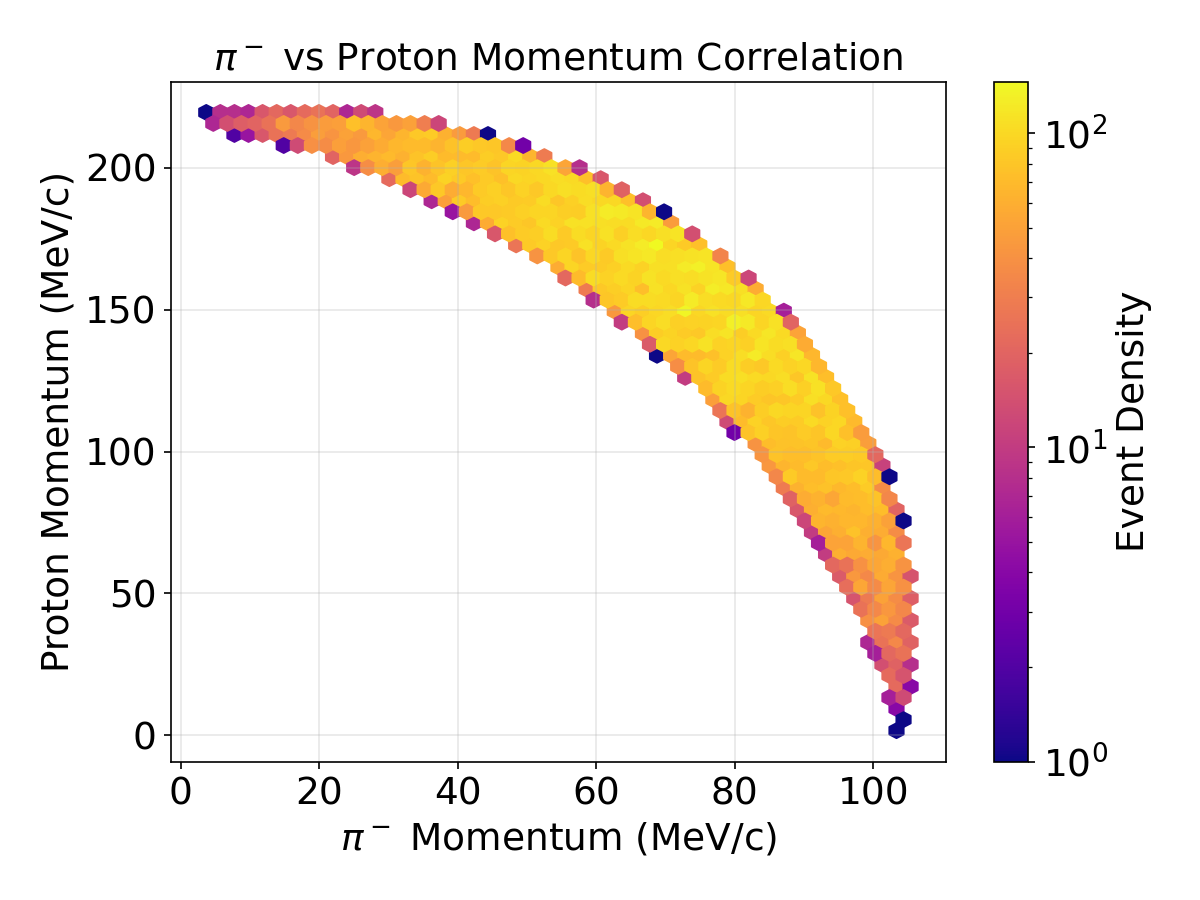}
        \caption{$\pi^-$ vs.\ proton}
        \label{fig:corr_pi_p}
    \end{subfigure}
    \hfill
    \begin{subfigure}[b]{0.48\textwidth}
        \centering
        \includegraphics[width=\textwidth]{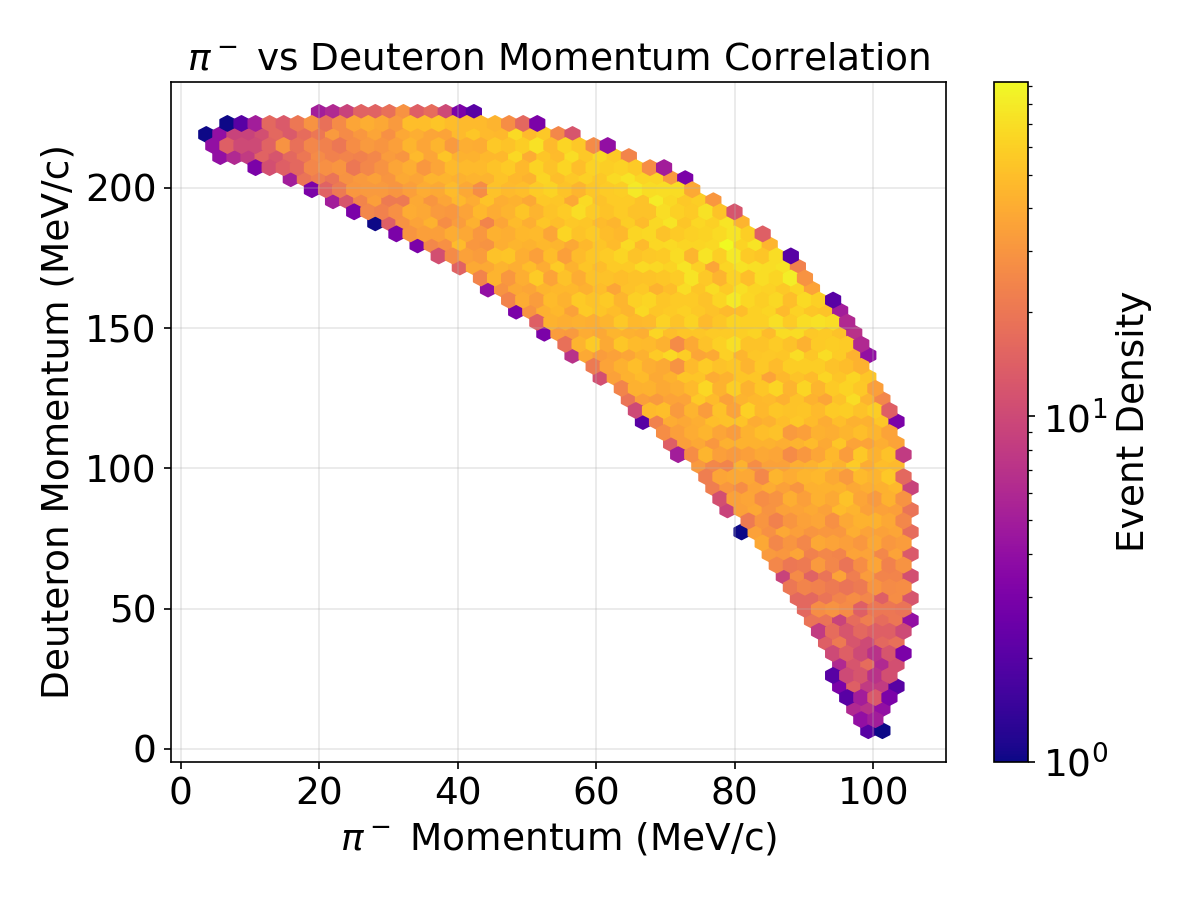}
        \caption{$\pi^-$ vs.\ deuteron}
        \label{fig:corr_pi_d}
    \end{subfigure}

    \vspace{1em}

    \begin{subfigure}[b]{0.48\textwidth}
        \centering
        \includegraphics[width=\textwidth]{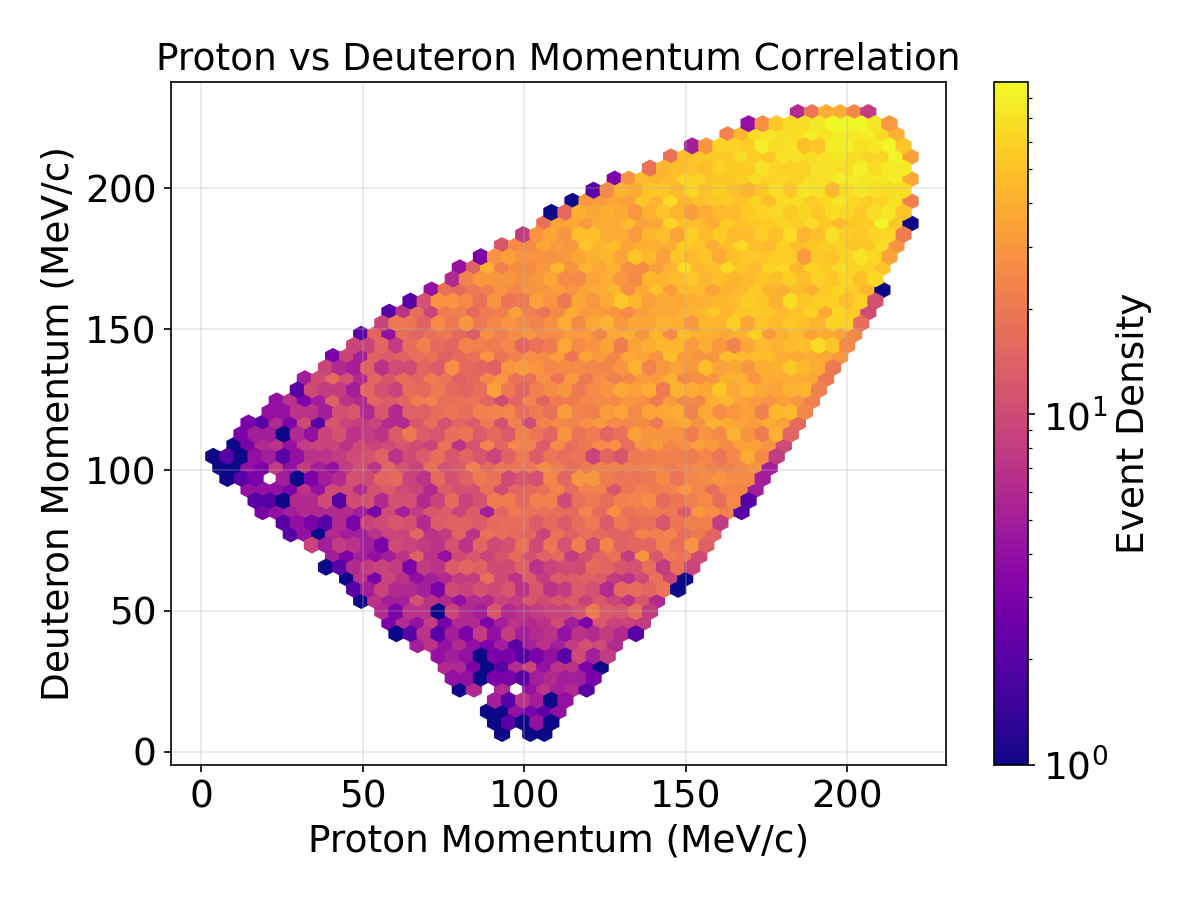}
        \caption{Proton vs.\ deuteron}
        \label{fig:corr_p_d}
    \end{subfigure}
    \hfill
    \begin{subfigure}[b]{0.48\textwidth}
        \centering
        \includegraphics[width=\textwidth]{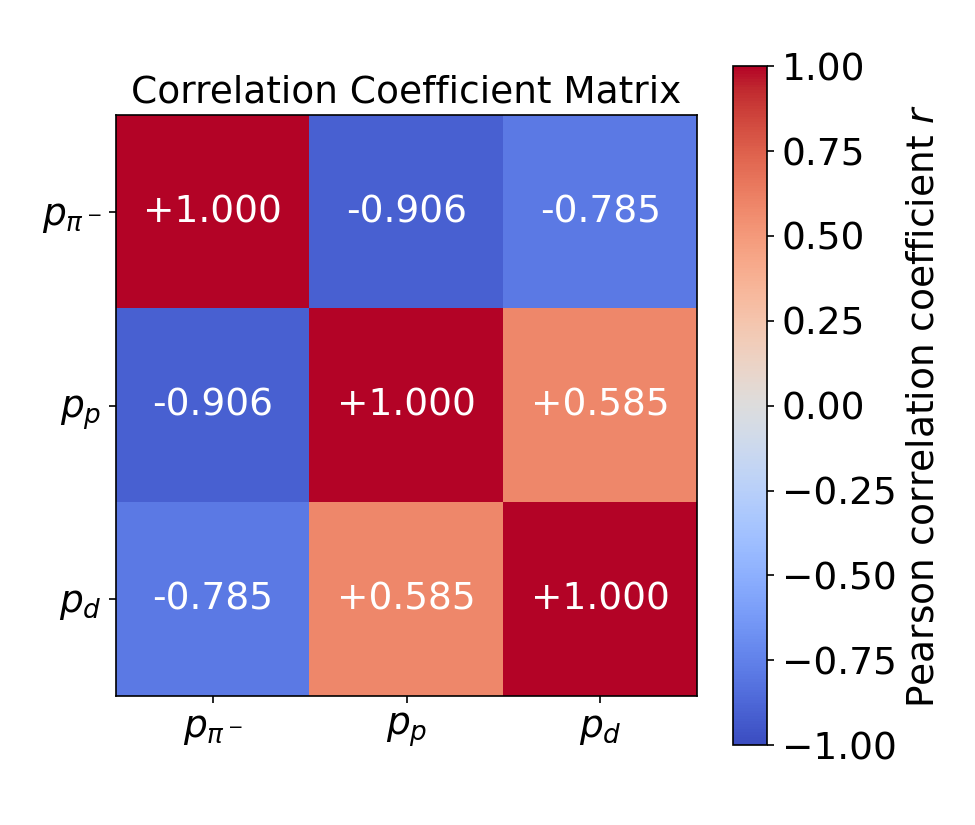}
        \caption{Pearson correlation coefficient matrix}
        \label{fig:corr_heatmap}
    \end{subfigure}

    \caption{Pairwise momentum correlations among the three decay products of
    $^3_\Lambda\text{H} \to d + p + \pi^-$. Panels (a)--(c) show hexbin density
    plots (logarithmic colour scale) of one particle's momentum magnitude
    against another's: (a) $\pi^-$ vs.\ proton, (b) $\pi^-$ vs.\ deuteron, and
    (c) proton vs.\ deuteron. The anti-correlated arcs reflect the kinematic anti-correlation between the momentum magnitudes of the pion and the nucleon/deuteron, arising from energy–momentum conservation. The broader, weakly correlated band in (c) reflects the more flexible relative kinematics between the proton and deuteron, once the pion momentum is fixed. Panel
    (d) summarizes these relationships as the Pearson correlation coefficient
    matrix among $p_{\pi^-}$, $p_p$, and $p_d$, computed over the unweighted
    phase-space sample: the pion momentum is strongly anti-correlated with
    both the proton ($r=-0.906$) and the deuteron ($r=-0.785$), while the
    proton and deuteron momenta are moderately positively correlated
    ($r=+0.585$), consistent with both nucleons sharing the recoil energy
    against the pion.}
    \label{fig:momentum_correlations}
\end{figure}

To characterise how the three decay-product momenta are kinematically
linked, we examine the pairwise correlations among $p_{\pi^-}$, $p_p$, and
$p_d$ across the unweighted phase-space sample. Figures~\ref{fig:corr_pi_p}--\ref{fig:corr_p_d} show the hexbin density distributions for each pair, and
Fig.~\ref{fig:corr_heatmap} summarizes the corresponding Pearson correlation
coefficients. The Pearson coefficients quantify the strength of the linear correlations induced by energy-momentum conservation.

The $\pi^-$--proton and $\pi^-$--deuteron correlations
(Figs.~\ref{fig:corr_pi_p}, \ref{fig:corr_pi_d}) both show a pronounced,
narrow, arc-like anti-correlation, with $r=-0.906$ and $r=-0.785$,
respectively. The $\pi^-$--proton pair reveals a tighter and more sharply defined arc compared to the visibly broader band for $\pi^-$--deuteron. By contrast, the proton--deuteron correlation (Fig.~\ref{fig:corr_p_d}) is broader and only moderately positively correlated, $r=+0.585$.

\subsubsection{Opening angle distributions}

\begin{figure}[htbp]
    \centering
    \begin{subfigure}[b]{0.48\textwidth}
        \centering
        \includegraphics[width=\textwidth]{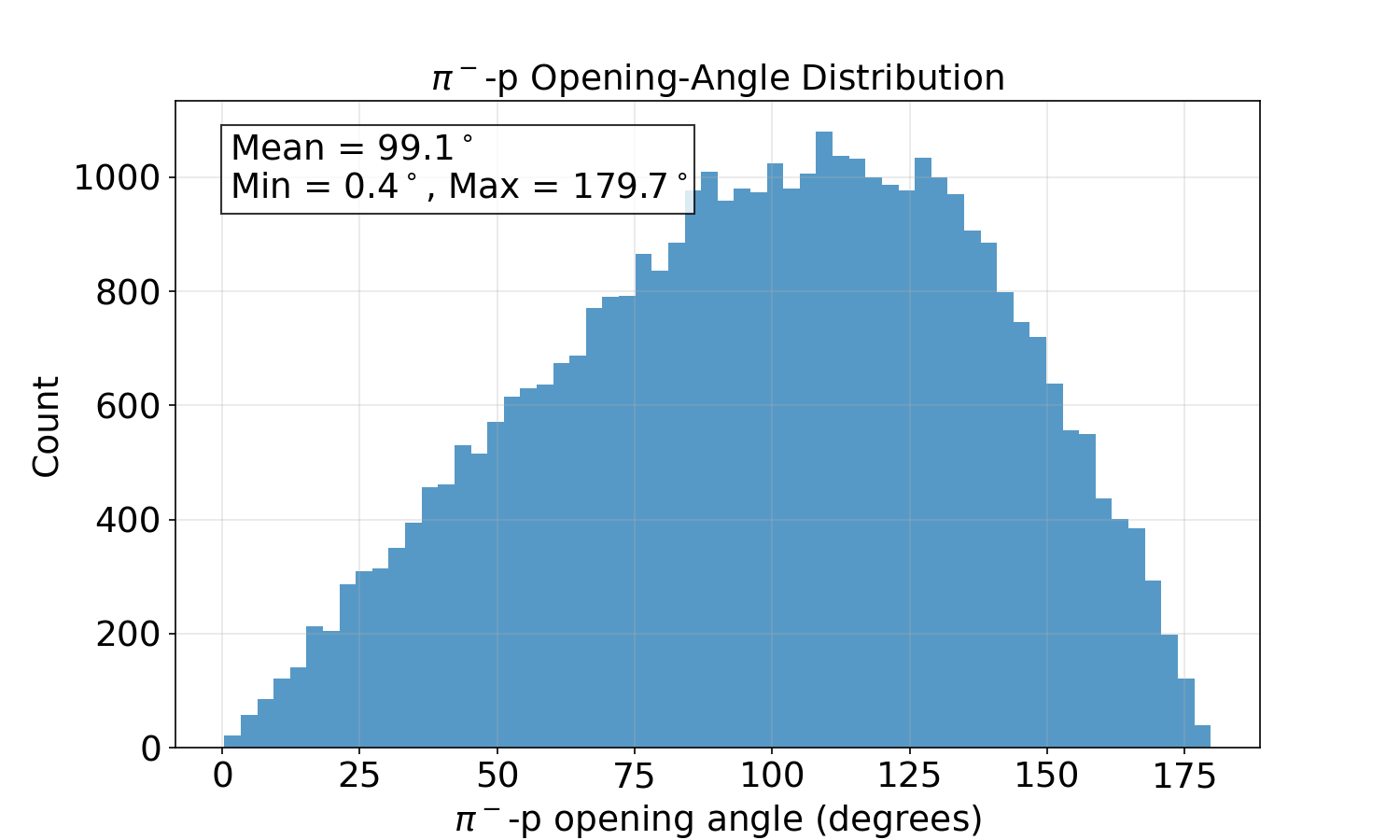}
        \caption{$\pi^-$--p}
        \label{fig:angle_pi_p}
    \end{subfigure}
    \hfill
    \begin{subfigure}[b]{0.48\textwidth}
        \centering
        \includegraphics[width=\textwidth]{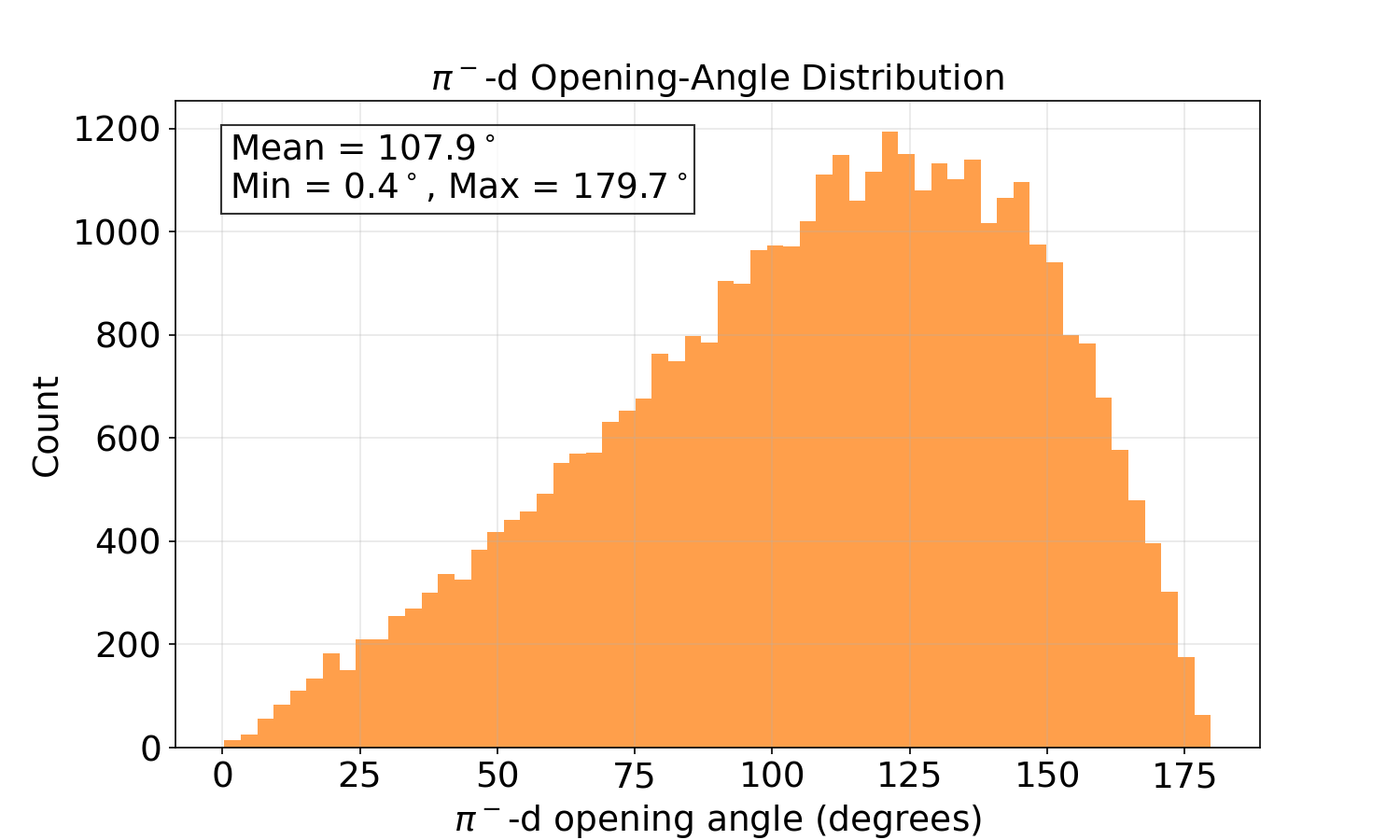}
        \caption{$\pi^-$--d}
        \label{fig:angle_pi_d}
    \end{subfigure}

    \vspace{1em}

    \begin{subfigure}[b]{0.48\textwidth}
        \centering
        \includegraphics[width=\textwidth]{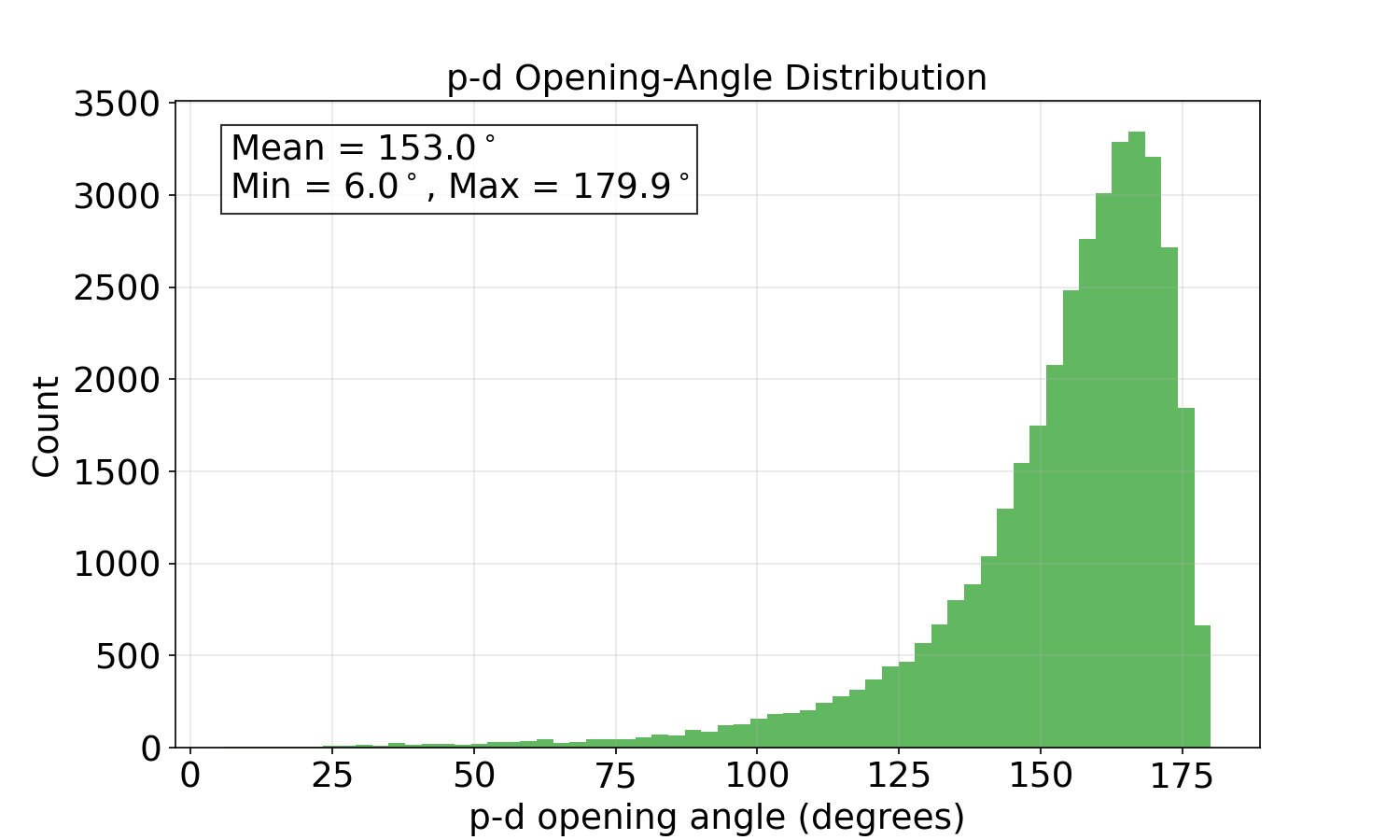}
        \caption{p--d}
        \label{fig:angle_p_d}
    \end{subfigure}

    \caption{Opening-angle distributions between pairs of decay products of
    $^3_\Lambda\text{H} \to d + p + \pi^-$, generated from unweighted
    phase-space events: (a) $\pi^-$--proton, (b) $\pi^-$--deuteron,
    and (c) proton--deuteron opening angles, all measured in the parent rest
    frame.}
    \label{fig:opening_angles}
\end{figure}

Figures~\ref{fig:angle_pi_p}--\ref{fig:angle_p_d} show the $\pi^-$--p,
$\pi^-$--d, and p--d opening angles (measured between each pair of outgoing
momenta in the parent rest frame) for the unweighted phase-space
sample. For a pair of decay products $i,j \in \{\pi^-, p, d\}$ with
three-momenta $\vec{p}_i$ and $\vec{p}_j$ in the hypertriton rest
frame, the opening angle $\theta_{ij}$ is the angle between
their momentum vectors at the instant of decay,

\begin{equation}
  \theta_{ij} \;=\; \arccos\!\left(
    \frac{\vec{p}_i \cdot \vec{p}_j}{|\vec{p}_i|\,|\vec{p}_j|}
  \right).
  \label{eq:opening_angle}
\end{equation}

This is computed event by event from the boosted final-state three-momenta of the unweighted, accepted events. Physically, $\theta_{ij}$ measures how collinear or back-to-back the
two particles emerge: $\theta_{ij} \to 0^\circ$ corresponds to the
pair moving in the same direction, while $\theta_{ij} \to 180^\circ$
corresponds to the pair emerging back-to-back. Because all three
final-state momenta lie in a common plane (from overall momentum
conservation, $\vec p_\pi + \vec p_p + \vec p_d = 0$), the three
pairwise opening angles $\theta_{\pi p}$, $\theta_{\pi d}$, and
$\theta_{pd}$ are not independent, and their distributions over the
generated event sample characterize the angular correlations imposed
by the three-body phase space.

The opening angle distributions in Figures~\ref{fig:angle_pi_p}--\ref{fig:angle_p_d} reveal a behaviour that mirrors the momentum correlations already discussed.

\begin{enumerate}
\item $\pi^-$--p and $\pi^-$--d opening angles: Both distributions are
broad, roughly symmetric around $90^\circ$--$110^\circ$. Also, they span the full
kinematically allowed range from near $0^\circ$ to near $180^\circ$. The
$\pi^-$--p distribution peaks near $99$--$110^\circ$ (mean $99.1^\circ$),
while the $\pi^-$--d distribution is shifted to somewhat larger angles,
peaking near $120$--$130^\circ$ (mean $107.9^\circ$). This shift reflects
the larger deuteron mass: since $M_d \approx 2 m_p$, the deuteron carries a
larger share of the recoil momentum against the pion on average, which
geometrically favours slightly larger opening angles relative to the lighter
proton. The wide spread in both distributions (essentially covering the
whole $0^\circ$--$180^\circ$ range) shows that the pion's direction
relative to the proton and to the deuteron is only loosely constrained. This is consistent with the three-body decay populating the full angular phase space rather than
favouring any fixed topology.

\item p--d opening angle: This distribution looks qualitatively
different: it is sharply peaked close to $180^\circ$ (mean $153.0^\circ$),
with almost no events below about $50^\circ$--$75^\circ$. This is the
clearest and most physically intuitive feature of the three plots: because the pion carries relatively little momentum over much of the phase-space sample, momentum conservation tends to make the proton and deuteron momenta approximately back-to-back. The pronounced peak just below $180^\circ$ reflects that small correction: the pion's own momentum tilts the p--d system slightly away from a perfectly collinear, back-to-back configuration.

\end{enumerate}

Together, these three distributions
paint a consistent picture of the decay topology: it resembles an
(approximately) two-body-like breakup of the recoiling $(p,d)$ system
against the light pion. The proton and deuteron are emitted nearly opposite each
other and the pion is free to point in a wide range of directions relative to
either one of them. As with the
momentum-correlation results, this angular structure is a pure
kinematic effect with no dynamical input; any future inclusion
of a final-state-interaction matrix element would be expected to
reshape these angular distributions relative to the pure phase-space
baseline shown here.

\subsubsection{Dalitz plots}

\begin{figure}[htbp]
    \centering
    \begin{subfigure}[b]{0.48\textwidth}
        \centering
        \includegraphics[width=\textwidth]{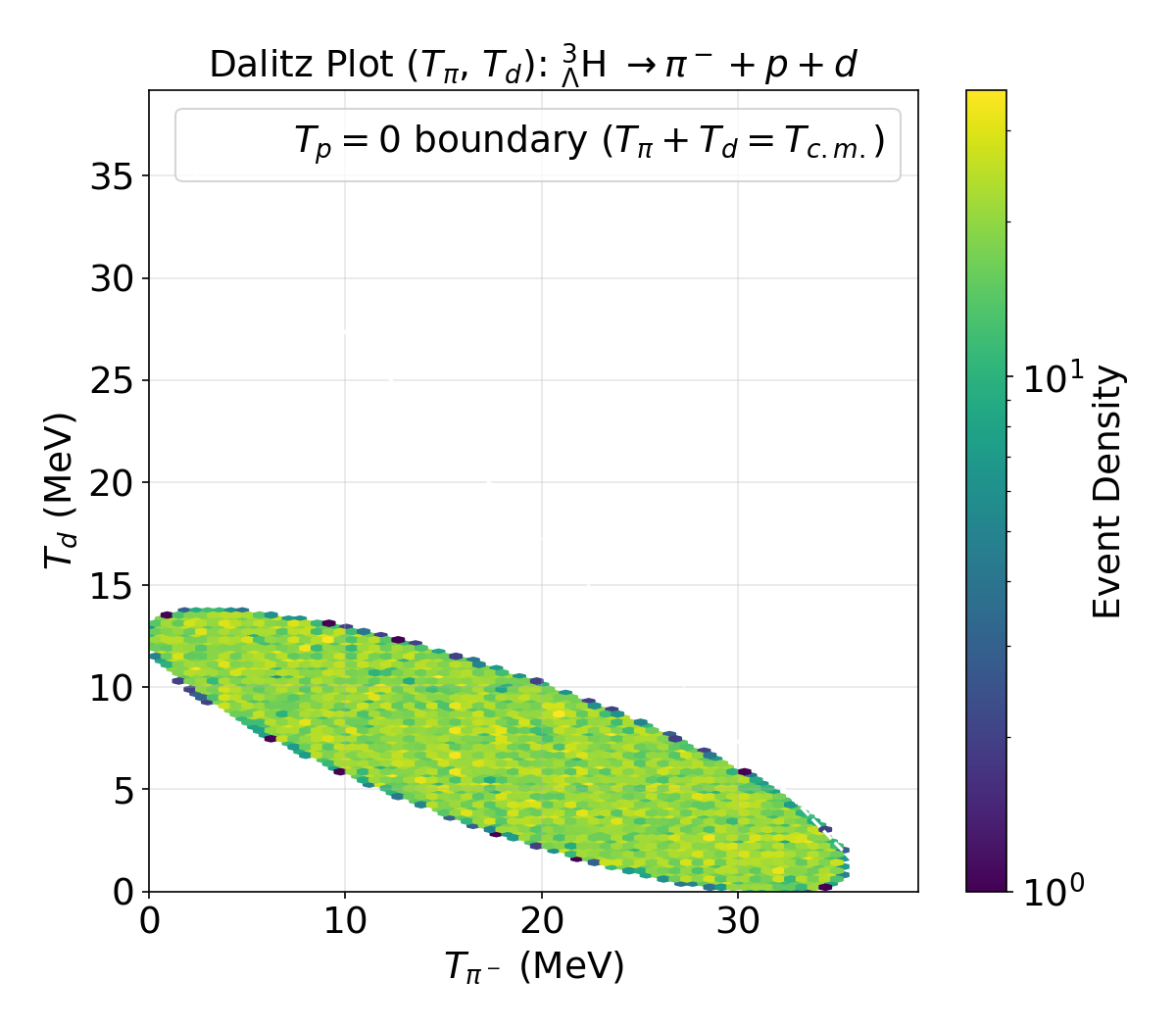}
        \caption{Kinetic-energy convention.}
        \label{fig:dalitz_kamada}
    \end{subfigure}
    \hfill
    \begin{subfigure}[b]{0.48\textwidth}
        \centering
        \includegraphics[width=\textwidth]{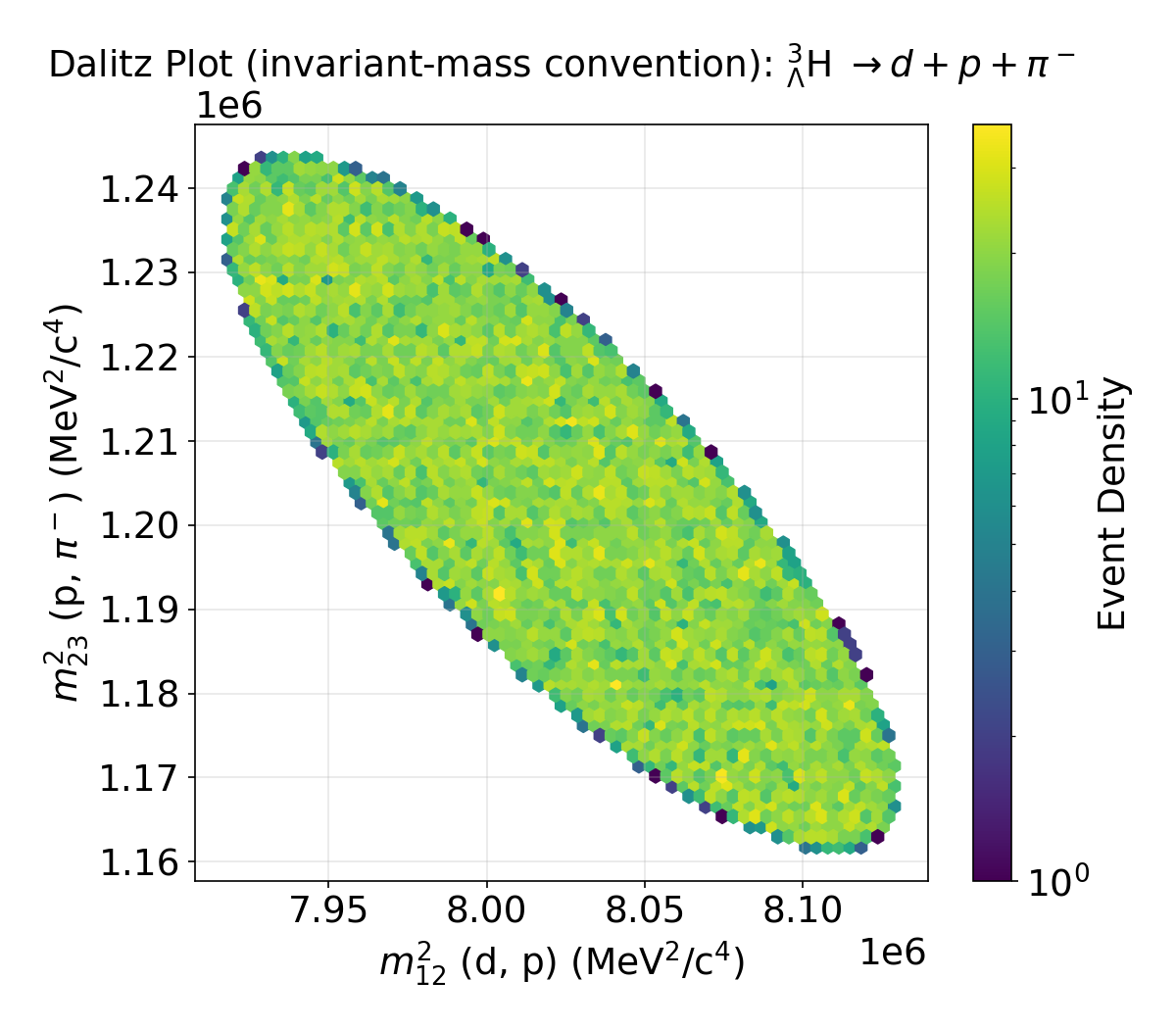}
        \caption{Relativistic invariant-mass-squared convention.}
        \label{fig:dalitz_invmass}
    \end{subfigure}
    \caption{Dalitz plots for the three-body decay $^3_\Lambda\text{H} \to
    d + p + \pi^-$, shown in two corresponding kinematic conventions. (a) The
    kinetic-energy convention of Ref \cite{kam1998},
    plotting $T_d$ vs.\ $T_{\pi^-}$. $T_{\pi^-}$ is relativistic kinetic energy while $T_d$ and $T_p$ are nonrelativistic. The kinetic energy ranges are $T_{\pi^-} \in [0.05, 35.32]$ MeV , $T_p \in [0.00, 25.69]$ MeV and $T_d \in [0.01, 13.72]$ MeV (b) The
    fully relativistic invariant-mass-squared convention, plotting
    $m^2_{23}(p,\pi^-)$ vs.\ $m^2_{12}(d,p)$. The ranges of the Dalitz variables are $m_{12}^2 (d,p)$: Min = 7918096.1, Max = 8129074.9 MeV$^2$/c$^4$ and $m_{23}^2 (p,\pi^- )$: Min = 1161777.4, Max = 1243520.8 MeV$^2$/c$^4$. In both conventions the population is uniform across the kinematically allowed region, as expected for pure (unweighted) three-body phase space with no dynamical matrix element applied.}
    \label{fig:dalitz_plots}
\end{figure}

Figures~\ref{fig:dalitz_kamada} and \ref{fig:dalitz_invmass} show two
corresponding representations of the three-body phase space populated by the
unweighted event sample for $^3_\Lambda\text{H} \to d + p + \pi^-$.

The Dalitz plot in Figure ~\ref{fig:dalitz_kamada} follows the same convention used in Ref. \cite{kam1998}, plotting the deuteron kinetic energy $T_d$ against the pion kinetic energy $T_{\pi^-}$, with the proton treated implicitly via energy conservation,
$T_{\pi^-}+T_p+T_d = T_{\text{c.m.}}$. The expressions for the relativistic pion kinetic energy $T_{\pi^-}$ and the nonrelativistic proton and deuteron kinetic energies are the same as in Ref. \cite{kam1998}, with our computed momenta used as input. This figure reveals that the population is uniform over the entire kinematically allowed region, bounded above by the $T_p=0$ line. This uniformity is the expected signature of unweighted phase space: with no dynamical matrix element applied, the phase-space density is uniform over the allowed region in the chosen Dalitz variables. The density simply reflects the available phase-space volume rather than any physical decay mechanism. In Ref. \cite{kam1998}, the "kinematically allowed area" is the analogue of our uniform region, but once the dynamics are folded in, essentially all of the decay strength piles up in a narrow region near $T_d \to 0$. The sharp contrast between a flat phase-space population and their strongly peaked,
low-$T_d$-dominated rate is itself the physical signature of interest. The comparison illustrates the extent to which dynamical effects, including the initial hypertriton wave function and final-state interactions, modify the purely kinematic phase-space distribution.

The relativistic invariant-mass-squared convention (Fig.~\ref{fig:dalitz_invmass}) presents the same information using $m^2_{23}(p,\pi^-)$ versus $m^2_{12}(d,p)$, the standard Lorentz-invariant
Dalitz variables. This convention has the advantage of being frame-independent and directly comparable across relativistic treatments, at the cost of a less intuitive kinetic-energy interpretation. As in the kinetic-energy convention, the density is uniform across the allowed (here
elongated, band-like) kinematic region, confirming that the sample correctly reproduces flat relativistic three-body phase space with no artifacts from the underlying $(T_{\pi^-}, T_d)$ parametrisation.

\subsection{Neutral pion mesonic decay}

While the negative-pion channel has been directly measured in heavy-ion collision experiments, and in earlier emulsion and bubble-chamber experiments, no experimental measurement of the neutral-pion three-body channel currently exists. The identical Raubold--Lynch sequential construction is applied to this neutral-pion three-body channel. The intermediate virtual system is now the neutron--deuteron subsystem \(X\), so the decay is factorised as

\begin{equation}
{}^3_\Lambda\mathrm{H}\to\pi^0+X,\qquad X\to n+d.
\end{equation}

The invariant mass \(m_{nd}\) is sampled uniformly inside the kinematic window

\begin{equation}
m_n+M_d\le m_{nd}\le M_0-m_{\pi^0},
\end{equation}
where \(M_0\) is the same as for the charged particle decay. The Q-value for the neutral pion decay is given by $Q_{\pi^0}^3= Q_{\pi^0}-B_{\Lambda}= 41.141 - 0.523 = 40.618 \text{ MeV} $. This allows more energy ($3.3$ MeV) for the decay products to share, compared to the case of the charged pion decay. The shift of the lower endpoint, $m_{nd}^{\min}-m_{pd}^{\min}=m_n-m_p\approx1.29$~MeV, follows from replacing the proton by the heavier neutron, since $M_d$ is common to both channels. The shift of the upper endpoint, $m_{nd}^{\max}-m_{pd}^{\max}=m_{\pi^-}-m_{\pi^0}\approx4.59$~MeV, follows from replacing the charged pion by the lighter neutral pion, with $M_0$ held fixed. The neutral-channel window is widened by $(m_{\pi^-}-m_{\pi^0})-(m_n-m_p)\approx3.30$~MeV, which is precisely the increase in the available three-body energy $Q^3_{\pi^0}-Q^3_{\pi^-}$.

Each event receives the phase-space weight formed from the two successive two-body breakup momenta. After the accept/reject step, the accepted events are distributed according to the normalized Lorentz-invariant three-body phase-space measure \(d\Phi_3\) for the \(\pi^0nd\) final state.

\begin{figure}[htbp]
    \centering
    \includegraphics[width=0.6\textwidth]{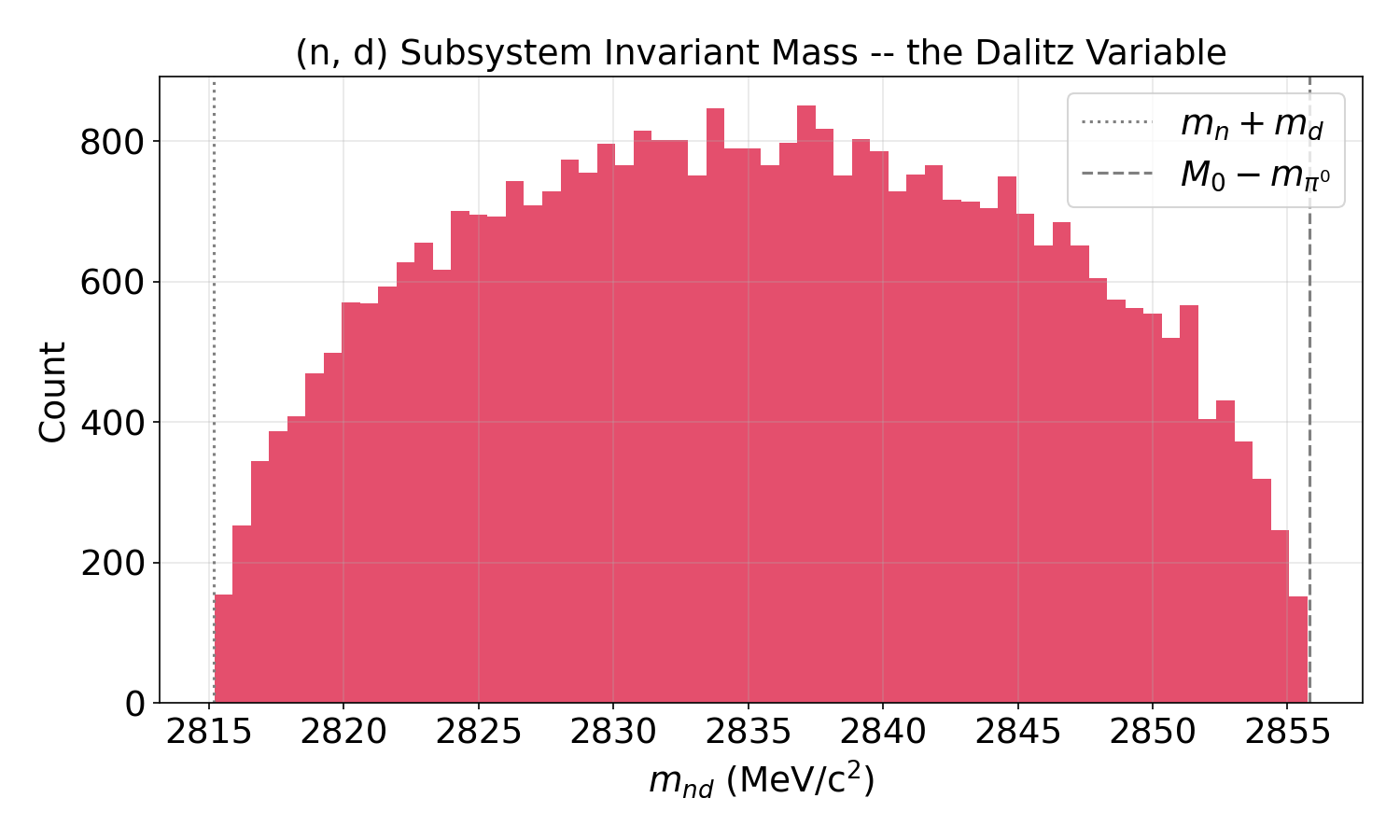}
    \caption{(n,d) subsystem invariant mass ($m_{nd}$) distribution. The minimum is 2815.21 and the maximum is 2855.74 MeV/$c^2$, with mean = 2835.36 MeV/$c^2$ and standard deviation = 10.16 MeV/$c^2$.}
    \label{fig:m_nd_invariant_mass}
\end{figure}

Figure~\ref{fig:m_nd_invariant_mass} shows the \(m_{nd}\) distribution. As in the charged channel, the differential phase-space density vanishes at both kinematic endpoints and reaches its maximum in the interior. The computed integrated volume \(\Phi_3^{\mathrm{tot}}\) is $2.569978 \times 10^{-1}$ MeV$^2$. 

\begin{figure}[htbp]
    \centering
    \begin{subfigure}[b]{0.48\textwidth}
        \centering
        \includegraphics[width=\textwidth]{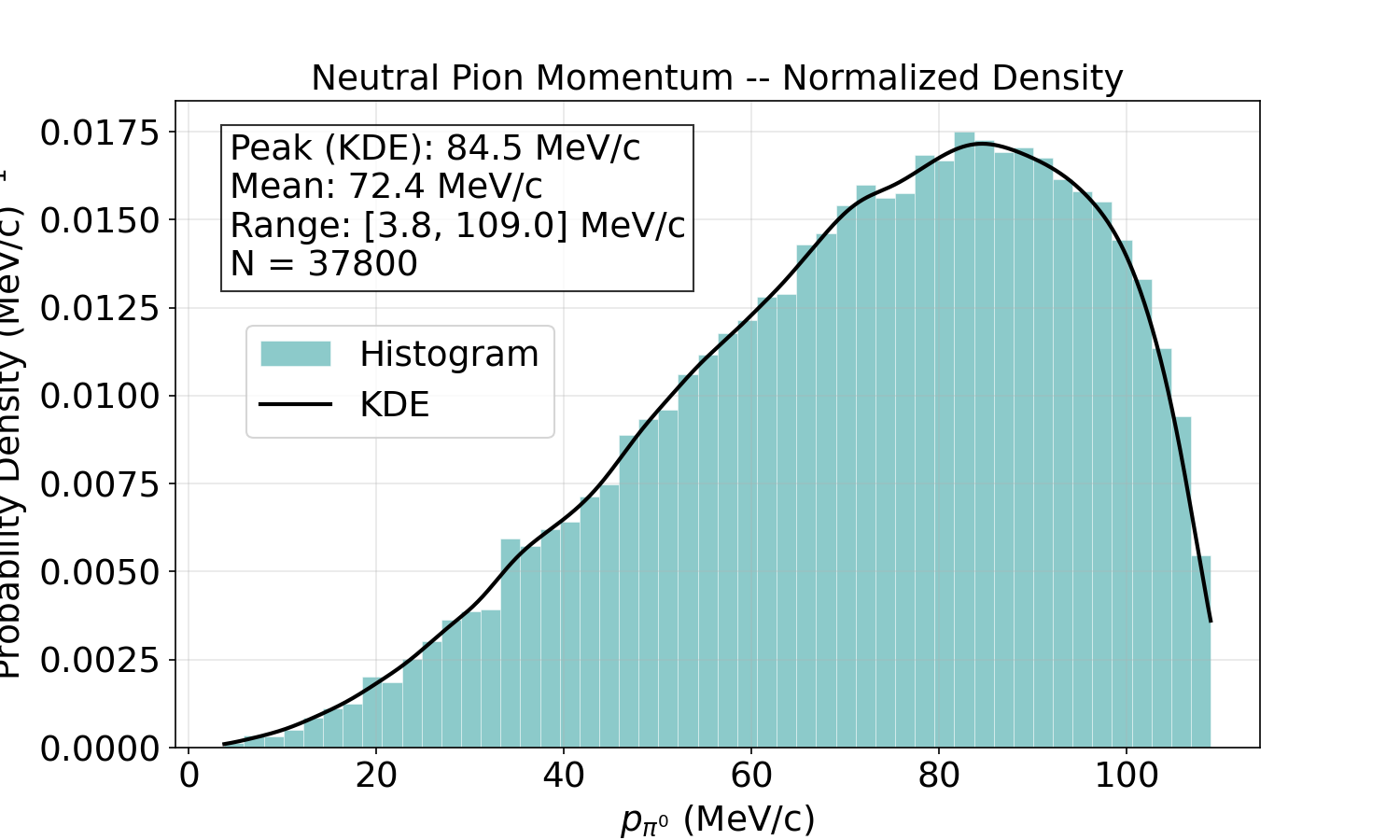}
        \caption{Pion momentum spectrum. Max = 108.95, mean = 72.38 MeV/c.}
        \label{fig:pion0_spectrum}
    \end{subfigure}
    \hfill
    \begin{subfigure}[b]{0.48\textwidth}
        \centering
        \includegraphics[width=\textwidth]{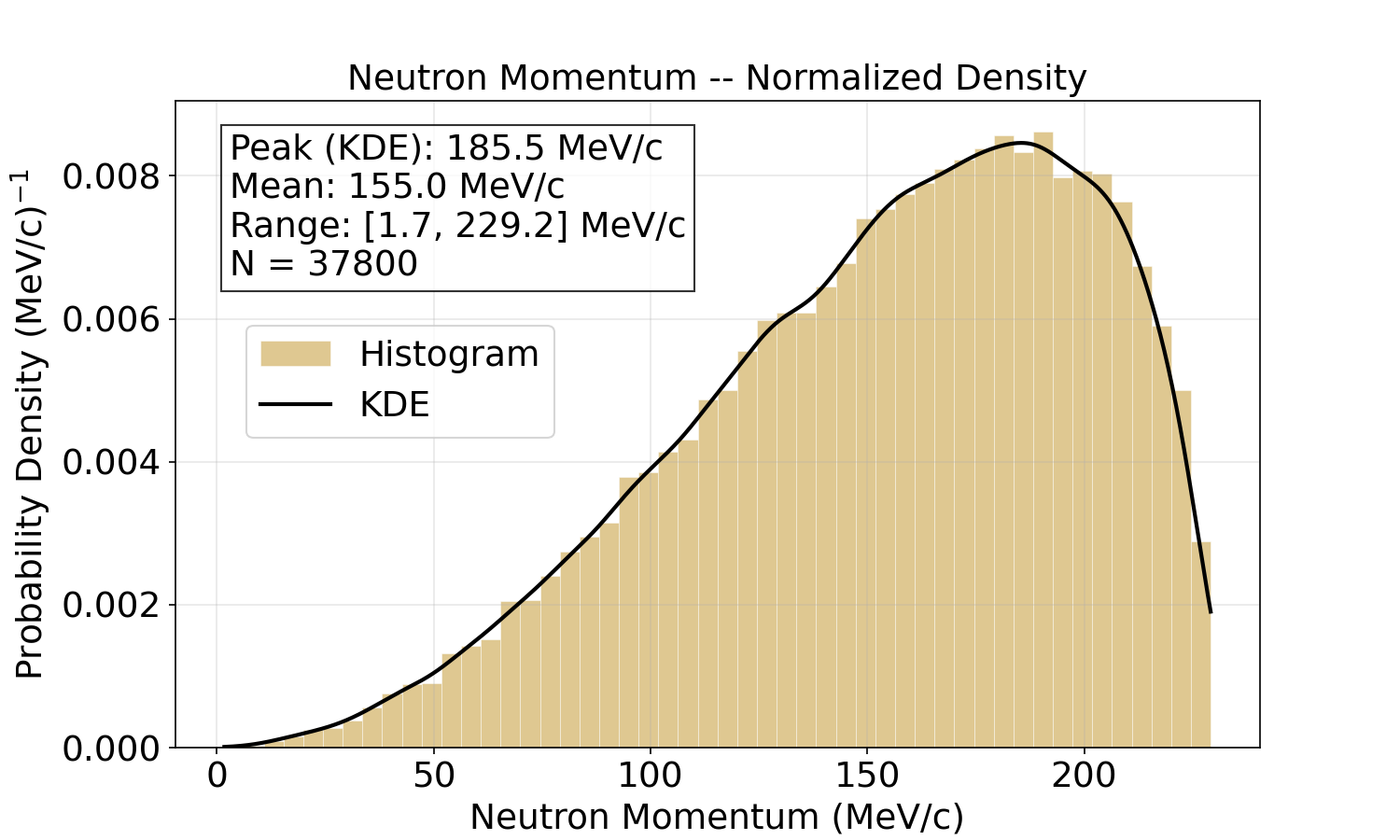}
        \caption{Neutron momentum spectrum. Max = 229.20, mean = 154.97 MeV/c}
        \label{fig:neutron_spectrum}
    \end{subfigure}
    \vspace{1em}
    \begin{subfigure}[b]{0.48\textwidth}
        \centering
        \includegraphics[width=\textwidth]{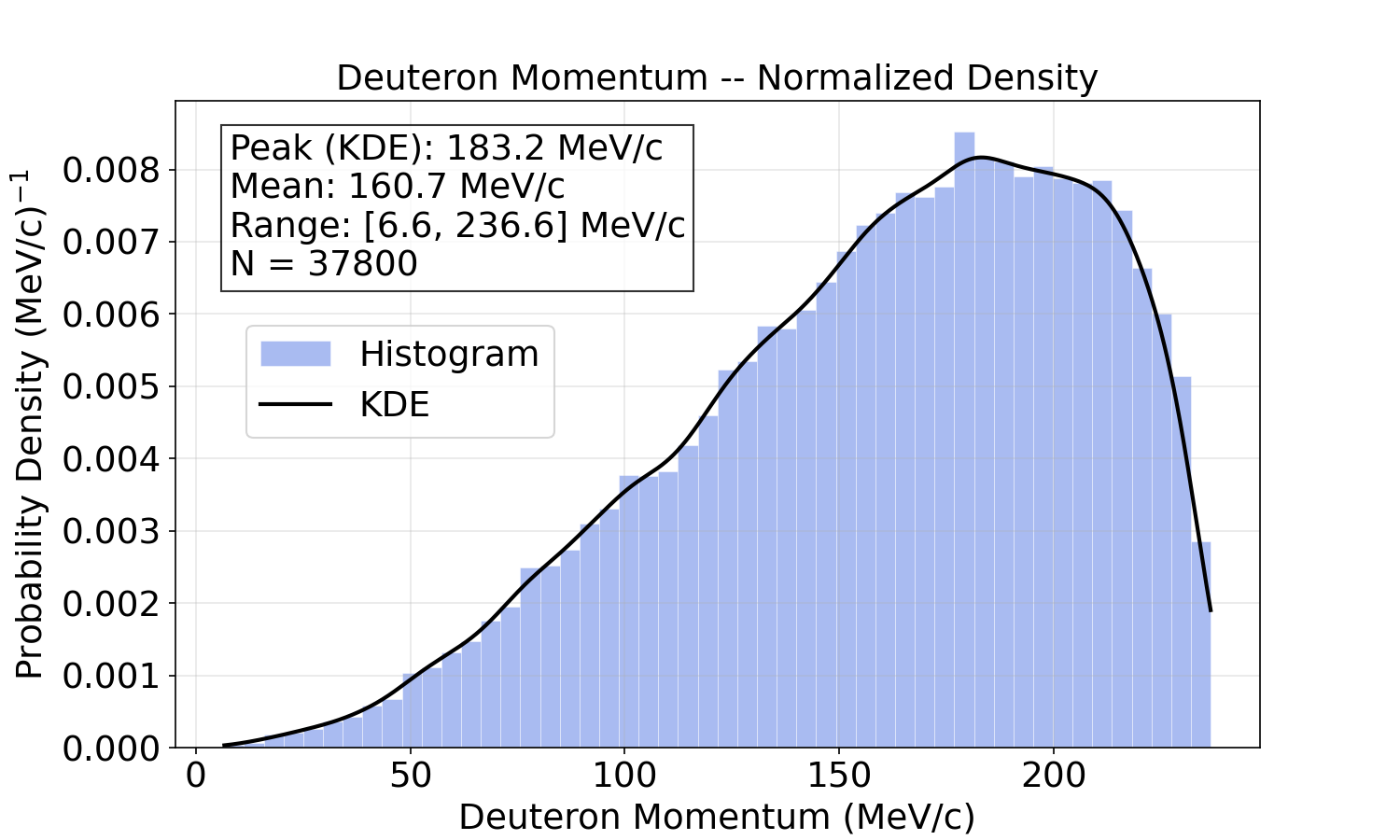}
        \caption{Deuteron momentum spectrum. Max = 236.61, mean = 160.66 MeV/c.}
        \label{fig:deuteron0_spectrum}
    \end{subfigure}
    \caption{Normalised momentum spectra for the three decay products of
    $^3_\Lambda\text{H} \to n + d + \pi^0$, generated using the
    Raubold--Lynch construction.}
    \label{fig:momentum_spectra_pi0}
\end{figure}

\subsubsection{Normalised probability density and histogram of individual momenta}

From $10^5$ raw weighted events, a total of 37800 accepted, unweighted events were obtained. The unweighted sample yields the momentum spectra of Figure \ref{fig:momentum_spectra_pi0}. The qualitative features are identical to those of the charged channel: long low-momentum tails, broad maxima and sharp kinematic cut-offs fixed by two-body recoil limits. Because \(m_n>m_p\) and \(m_{\pi^0}<m_{\pi^-}\) (i.e. more energy is available to the decay products), the numerical locations of peaks, means and endpoints are all shifted to higher values.

The momentum of the two-body mesonic decay was found to be $p_{\pi^0}=118.129  $ MeV/c. This monochromatic momentum of the two-body decay is a theoretical prediction obtained from the root-finding framework of Ref. \cite{meo2026b}. Unlike the charged channel, no experimental measurement of any neutral-channel observable currently exists. This predicted two-body momentum lies about 9.179 MeV/c above the maximum pion momentum in the three-body continuum computed here (108.95 MeV/c). Therefore, the two endpoints are well-separated on kinematic grounds.

\begin{figure}[htbp]
    \centering
    \begin{subfigure}[b]{0.48\textwidth}
        \centering
        \includegraphics[width=\textwidth]{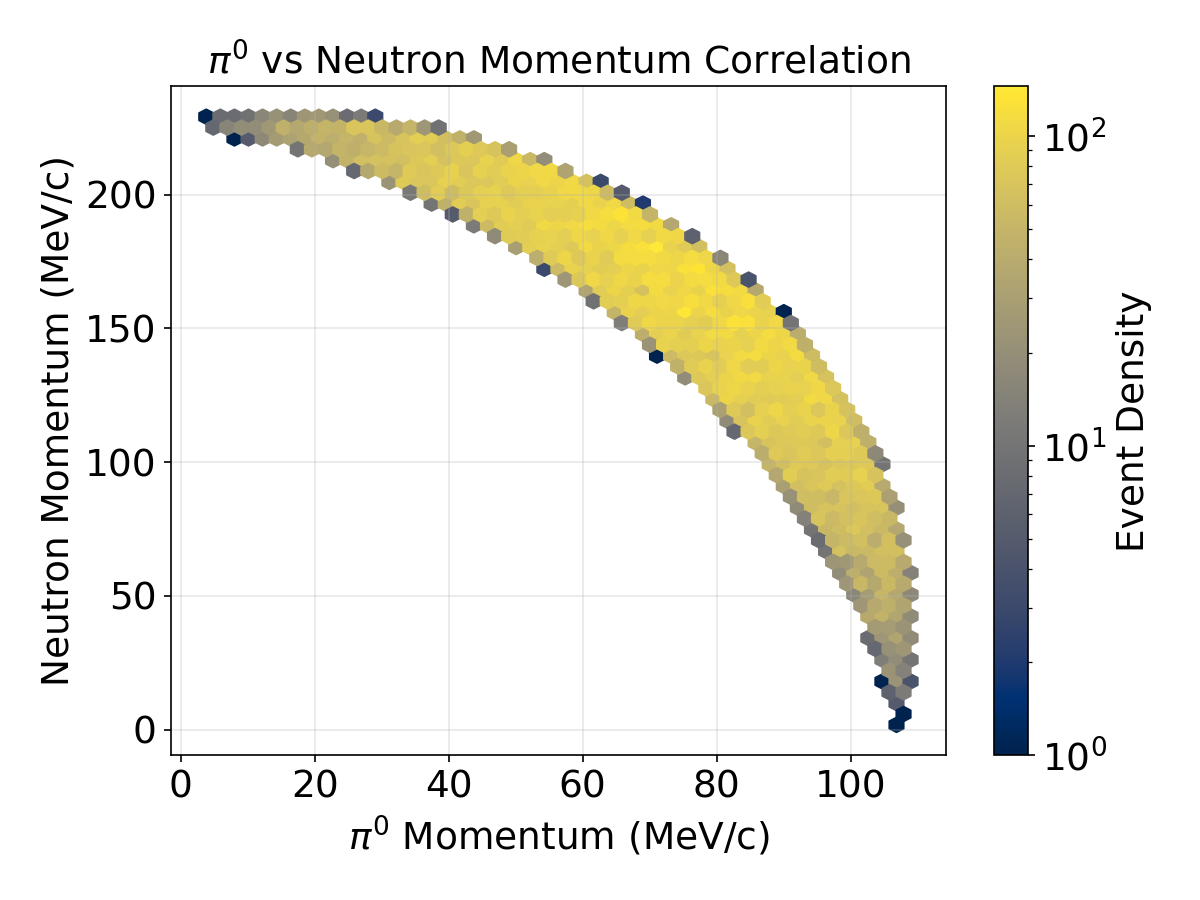}
        \caption{$\pi^0$ vs.\ neutron}
        \label{fig:corr_pi0_n}
    \end{subfigure}
    \hfill
    \begin{subfigure}[b]{0.48\textwidth}
        \centering
        \includegraphics[width=\textwidth]{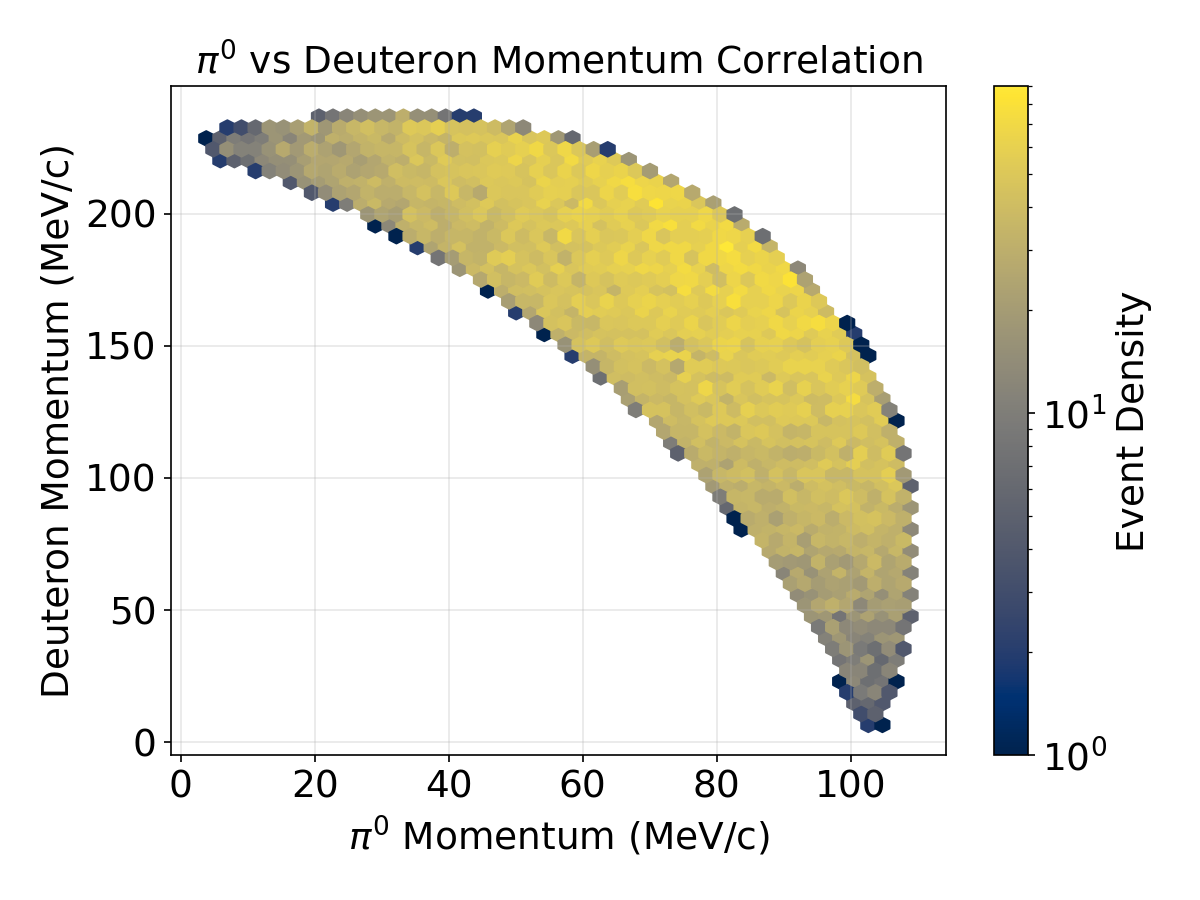}
        \caption{$\pi^0$ vs.\ deuteron}
        \label{fig:corr_pi0_d}
    \end{subfigure}
    \vspace{1em}
    \begin{subfigure}[b]{0.48\textwidth}
        \centering
        \includegraphics[width=\textwidth]{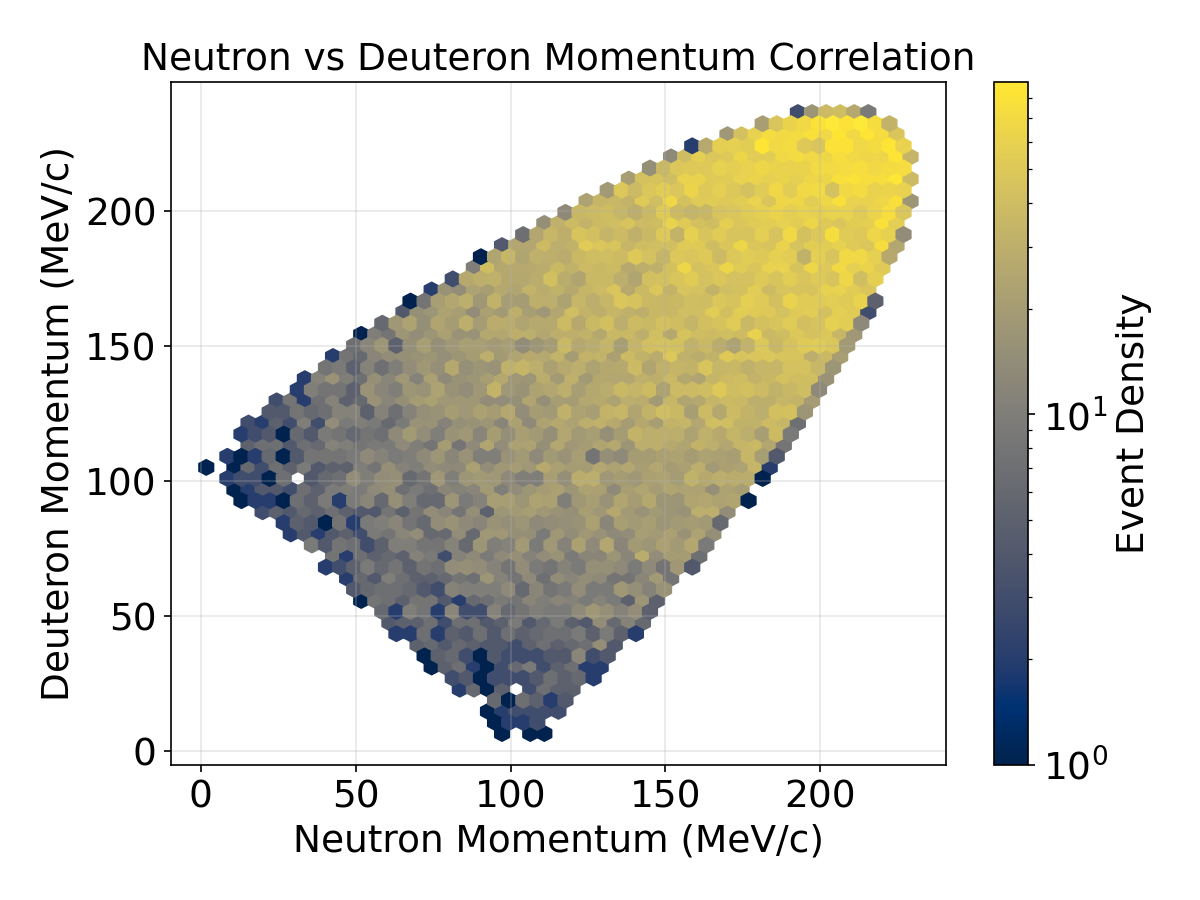}
        \caption{Neutron vs.\ deuteron}
        \label{fig:corr_n_d}
    \end{subfigure}
    \hfill
    \begin{subfigure}[b]{0.48\textwidth}
        \centering
        \includegraphics[width=\textwidth]{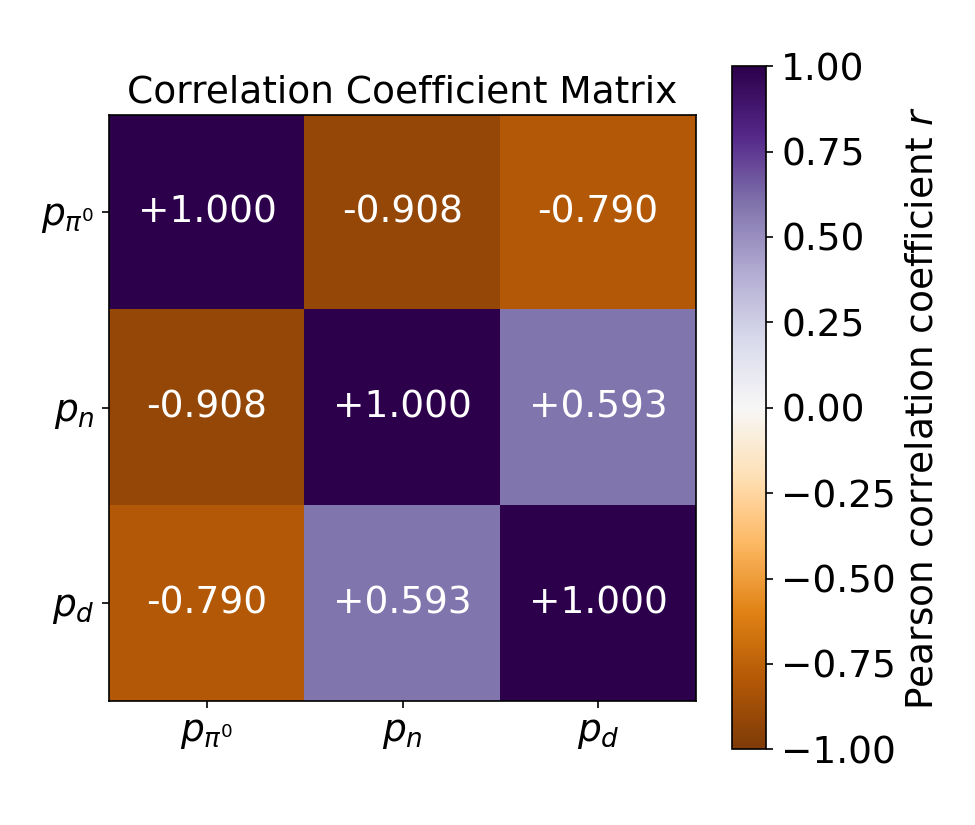}
        \caption{Pearson correlation coefficient matrix}
        \label{fig:corr_heatmap_pi0}
    \end{subfigure}
    \caption{Pairwise momentum correlations among the three decay products of
    $^3_\Lambda\text{H} \to d + n + \pi^0$. Panels (a)--(c) show hexbin density
    plots (logarithmic colour scale) of one particle's momentum magnitude
    against another's: (a) $\pi^0$ vs.\ neutron, (b) $\pi^0$ vs.\ deuteron, and
    (c) neutron vs.\ deuteron. The arcs reflect the kinematic anti-correlation between the momentum magnitudes of the pion and the nucleon/deuteron, arising from energy–momentum conservation. The broader, weakly
    correlated band in (c) reflects the more flexible relative kinematics
    between the neutron and deuteron, once the pion momentum is fixed. Panel
    (d) summarizes these relationships as the Pearson correlation coefficient
    matrix among $p_{\pi^0}$, $p_n$, and $p_d$, computed over the unweighted
    phase-space sample: the pion momentum is strongly anti-correlated with
    both the neutron ($r=-0.908$) and the deuteron ($r=-0.790$), while the
    neutron and deuteron momenta are moderately positively correlated
    ($r=+0.593$), consistent with the neutron and deuteron sharing the recoil energy
    against the pion.}
    \label{fig:momentum_correlations_pi0}
\end{figure}

\subsubsection{Pairwise momentum correlation between decay products}

Figure~\ref{fig:momentum_correlations_pi0} displays the same pattern already observed for the charged channel: strong arc-like anti-correlations of the pion with the neutron and with the deuteron, and a milder positive correlation between the neutron and deuteron. These are pure kinematic effects of three-body momentum conservation. The corresponding Pearson correlation coefficients are similar to those obtained for the charged-pion channel.

\subsubsection{Opening angle distributions}

\begin{figure}[htbp]
    \centering
    \begin{subfigure}[b]{0.48\textwidth}
        \centering
        \includegraphics[width=\textwidth]{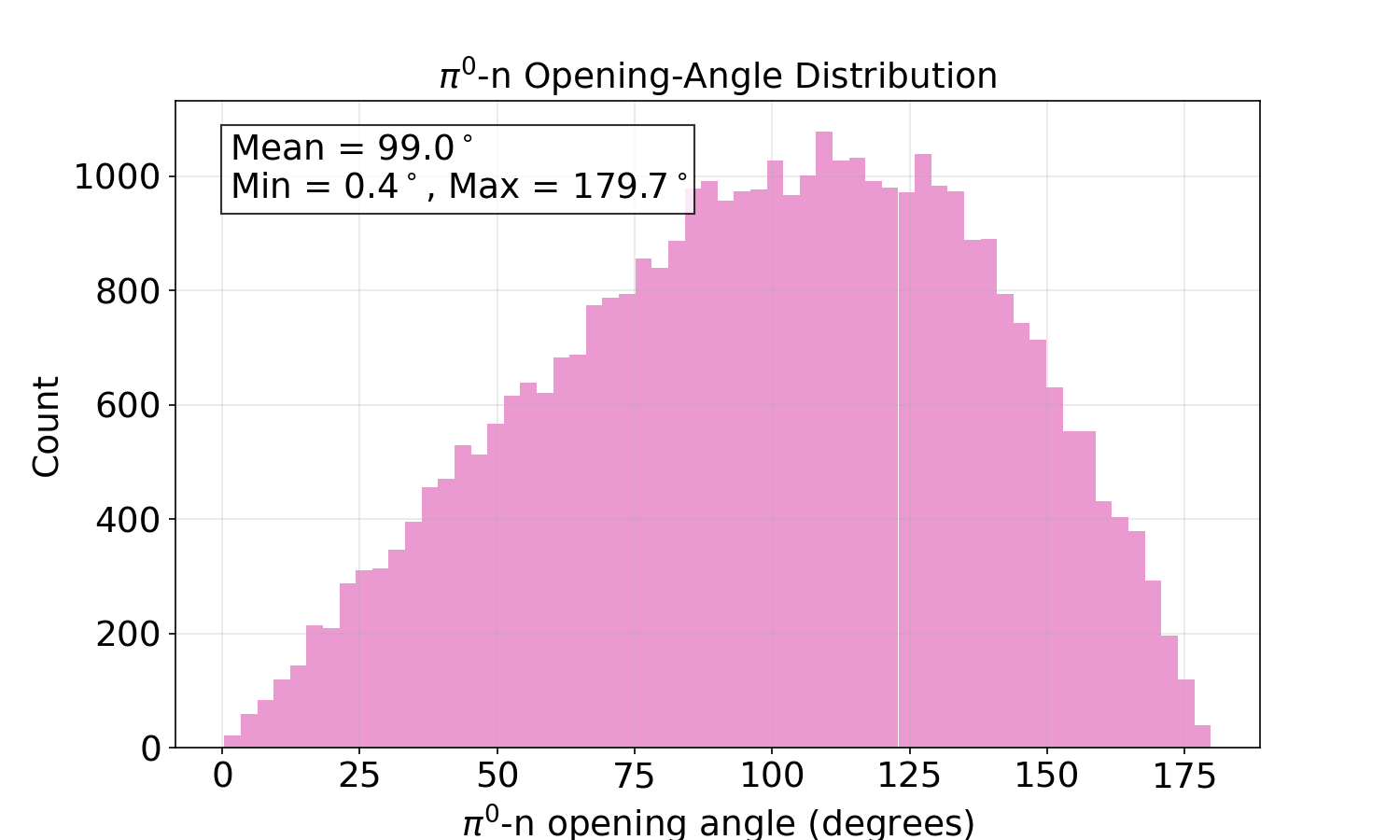}
        \caption{$\pi^0$--n}
        \label{fig:angle_pi0_n}
    \end{subfigure}
    \hfill
    \begin{subfigure}[b]{0.48\textwidth}
        \centering
        \includegraphics[width=\textwidth]{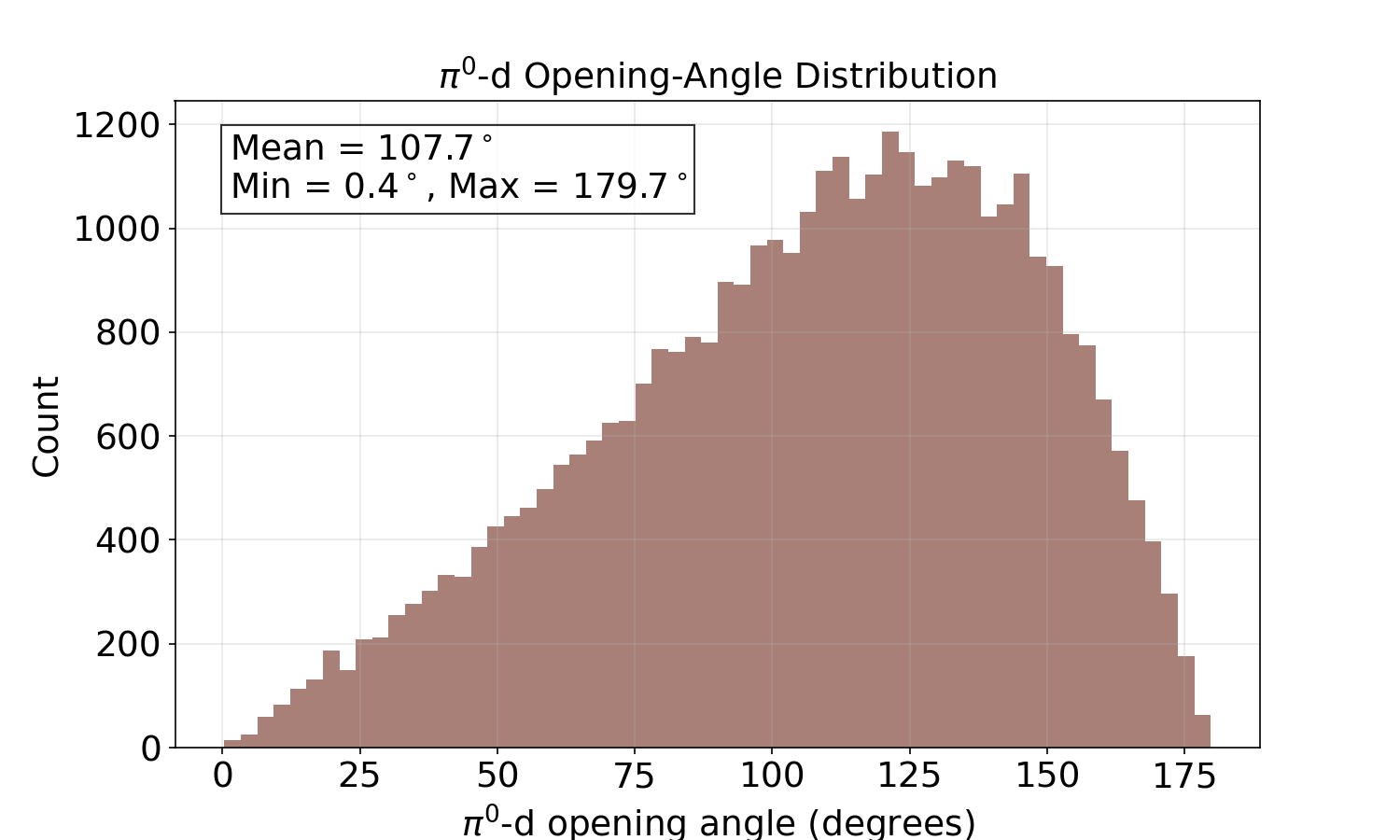}
        \caption{$\pi^0$--d}
        \label{fig:angle_pi0_d}
    \end{subfigure}
    \vspace{1em}
    \begin{subfigure}[b]{0.48\textwidth}
        \centering
        \includegraphics[width=\textwidth]{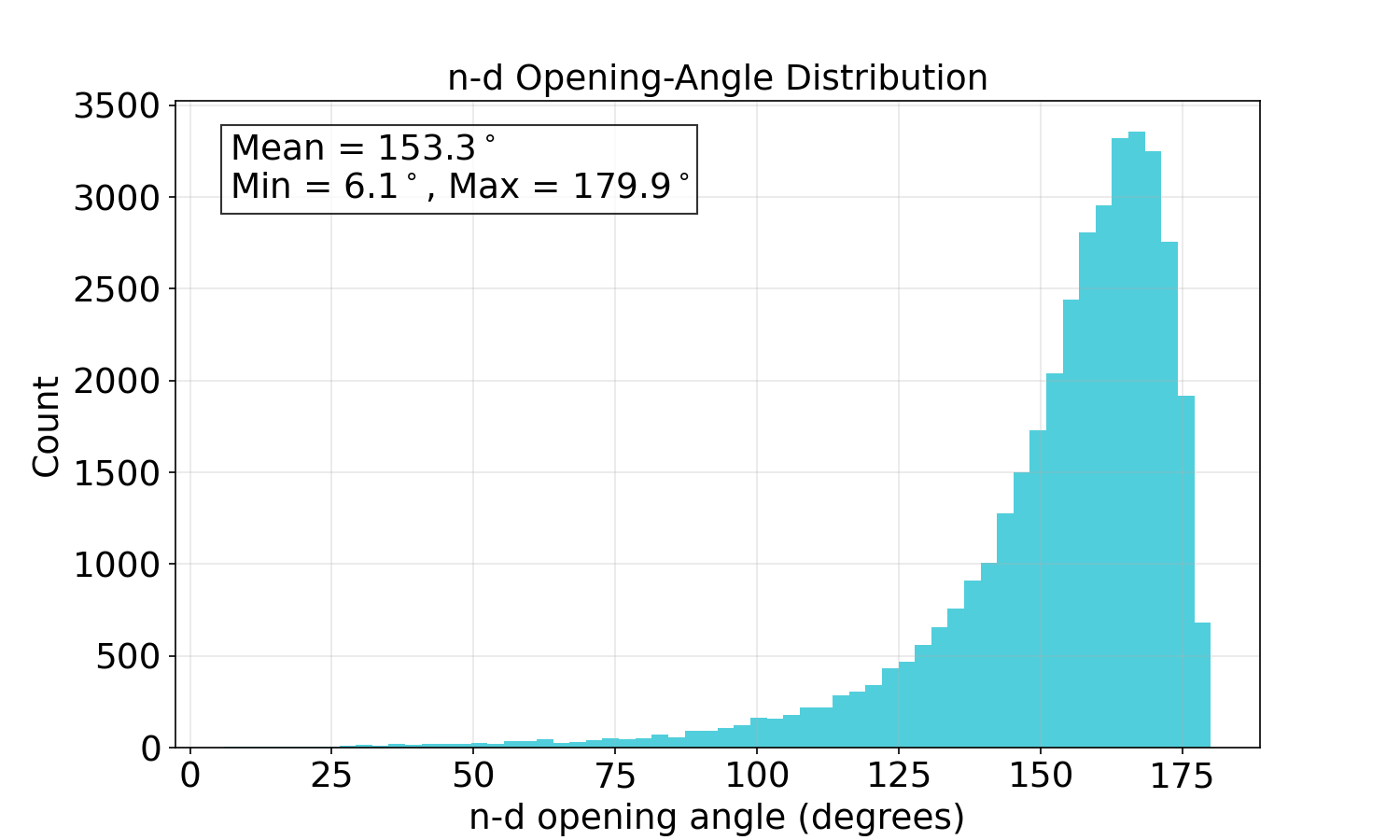}
        \caption{n--d}
        \label{fig:angle_n_d}
    \end{subfigure}
    \caption{Opening-angle distributions between pairs of decay products of
    $^3_\Lambda\text{H} \to d + n + \pi^0$, generated from unweighted
    phase-space events: (a) $\pi^0$--neutron, (b) $\pi^0$--deuteron,
    and (c) neutron--deuteron opening angles, all measured in the parent rest
    frame.}
    \label{fig:opening_angles_pi0}
\end{figure}

The opening-angle distributions of Figure \ref{fig:opening_angles_pi0}
 mirror those of the charged channel. The \(\pi^0\)--\(n\) and \(\pi^0\)--\(d\) spectra are broad and roughly centred near \(90^\circ\)--\(110^\circ\), while the \(n\)--\(d\) spectrum is sharply peaked close to \(180^\circ\). The topology again resembles an approximately two-body-like breakup of the recoiling \((n,d)\) system against the light pion.

\subsubsection{Dalitz plots}

\begin{figure}[htbp]
    \centering
    \begin{subfigure}[b]{0.48\textwidth}
        \centering
        \includegraphics[width=\textwidth]{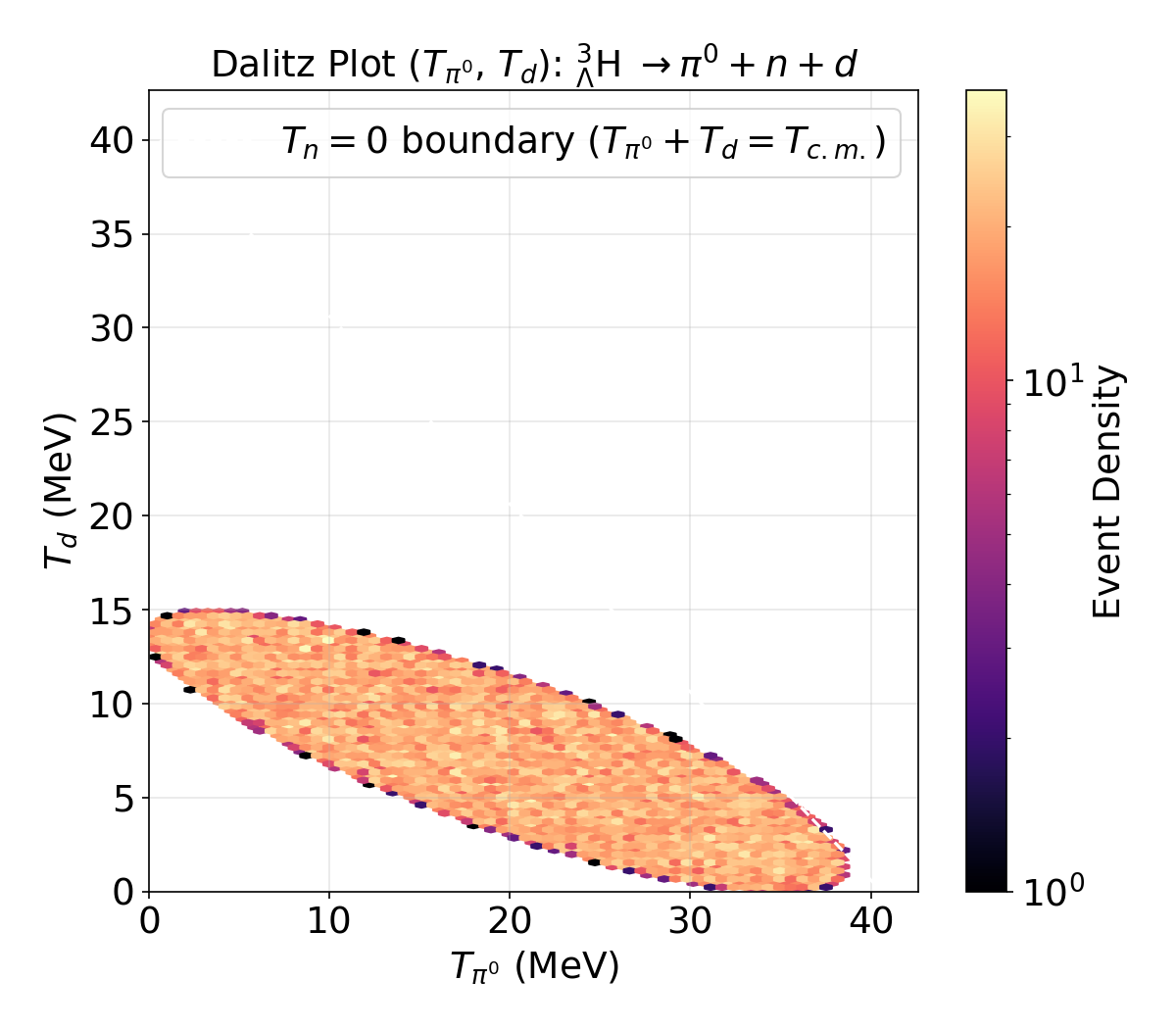}
        \caption{Kinetic-energy convention.}
        \label{fig:dalitz_kamada_pi0}
    \end{subfigure}
    \hfill
    \begin{subfigure}[b]{0.48\textwidth}
        \centering
        \includegraphics[width=\textwidth]{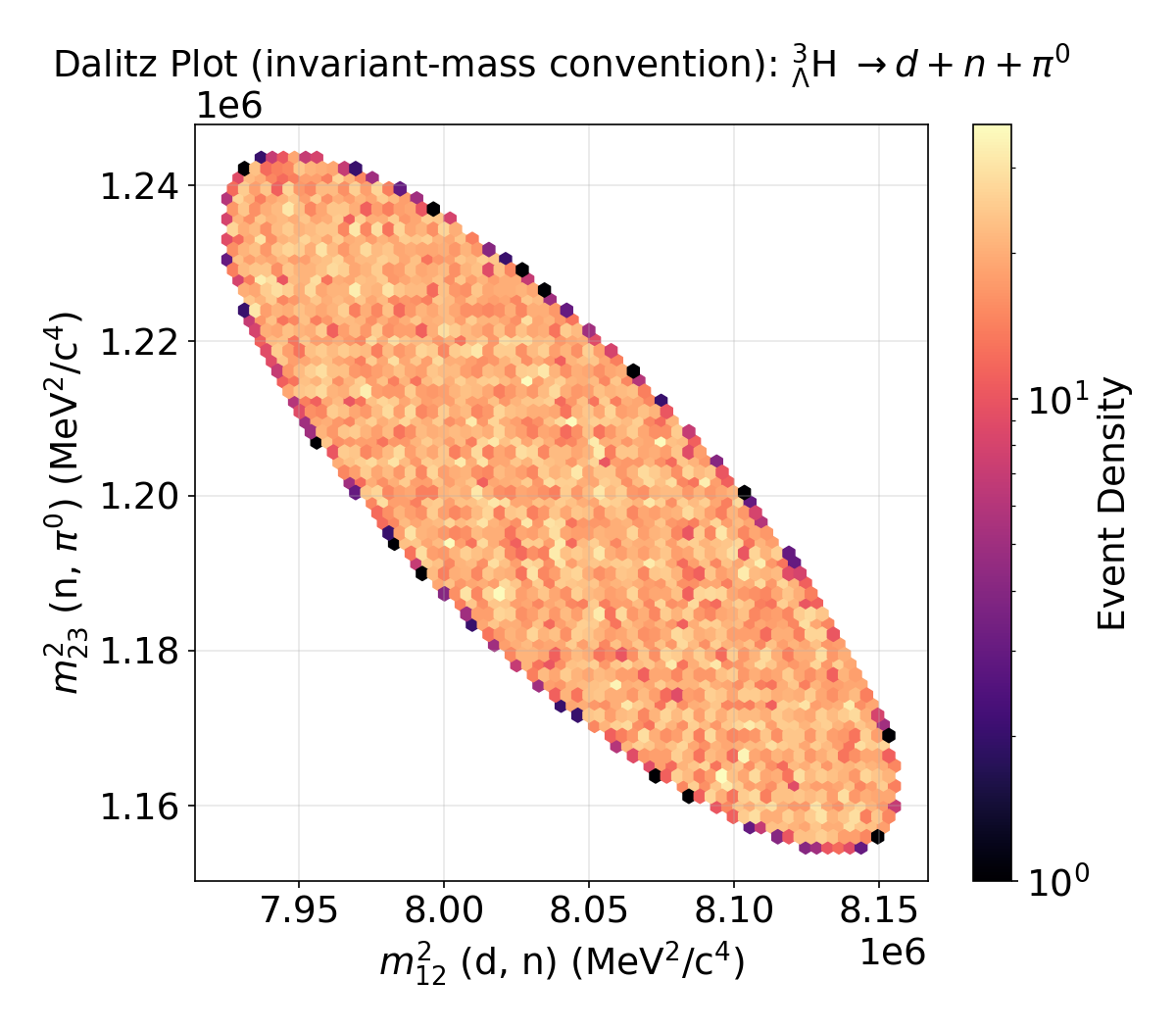}
        \caption{Relativistic invariant-mass-squared convention.}
        \label{fig:dalitz_invmass_pi0}
    \end{subfigure}
    \caption{Dalitz plots for the three-body decay $^3_\Lambda\text{H} \to
    d + n + \pi^0$, shown in two corresponding kinematic conventions. (a) The
    kinetic-energy convention of Ref \cite{kam1998},
    plotting $T_d$ vs.\ $T_{\pi^0}$. $T_{\pi^0}$ is relativistic kinetic energy while $T_d$ and $T_n$ are nonrelativistic. The kinetic energy ranges are $T_{\pi^0} \in [0.05, 38.48]$ MeV , $T_n \in [0.00, 27.55]$ MeV and $T_d \in [0.01, 14.90]$ MeV (b) The
    fully relativistic invariant-mass-squared convention, plotting
    $m^2_{23}(n,\pi^0)$ vs.\ $m^2_{12}(d,n)$. The Dalitz variables have the ranges $m_{12}^2 (d,n)$: Min = 7925387.6, Max = 8155264.0 MeV$^2$/c$^4$ and $m_{23}(n,pi0)$ : Min = 1154671.4, Max = 1243517.2 MeV$^2$/c$^4$. In both conventions the
    population is uniform across the kinematically allowed region, as
    expected for pure (unweighted) three-body phase space with no
    dynamical matrix element applied.}
    \label{fig:dalitz_plots_pi0}
\end{figure}

Figure \ref{fig:dalitz_plots_pi0} shows the two standard Dalitz representations. As in the charged channel, the population is uniform across the entire kinematically allowed region in both the kinetic-energy and invariant-mass-squared conventions. This is the pure-phase-space baseline against which any future dynamical matrix element can be compared.

\subsubsection{Monte Carlo integration over phase space.}

The unweighted sample generated for the neutral channel already carries the exact Lorentz-invariant phase-space measure \(d\Phi_3\). Consequently the same Monte-Carlo estimator developed earlier applies. Once a spin-averaged squared matrix element \(|\mathcal{M}(p_{\pi^0},p_n,p_d)|^2\) is available, the partial decay width is obtained as

\begin{equation}
\Gamma_{\pi^0}^3=\frac{\Phi_3^{\mathrm{tot}}}{2M_0}\frac1N\sum_{i=1}^N|\mathcal{M}_i|^2,
\end{equation}

where \(\Phi_3^{\mathrm{tot}}\) is the total three-body phase-space volume evaluated for the \(\pi^0nd\) final state and the sum runs over the accepted events. No regeneration of the kinematics is required; any dynamical input can be incorporated through event-by event matrix-element weighting of the existing phase-space sample. This cleanly separates the phase-space construction of the present work from subsequent dynamical calculations.

\newpage
\section{Conclusion}

The Raubold–Lynch phase-space generator has been applied to the three-body mesonic decay of the hypertriton in both the charged and neutral pion channels. The method exploits the recursive factorisation of the Lorentz-invariant three-body phase space into two exact two-body decays through an intermediate virtual nucleon–deuteron subsystem. This yields complete final-state four-momenta for each Monte Carlo event with 4–momentum conservation.

From the accepted, unweighted events generated, momentum spectra, pairwise correlations, opening angles, and Dalitz plots were constructed. These kinematic distributions exhibit the characteristic features of pure three-body phase space, including endpoint suppression in the subsystem invariant-mass spectra and the pion recoils back-to-back against the total momentum of the nucleon–deuteron subsystem. Furthermore, these distributions provide a transparent, model-independent reference for identifying and quantifying dynamical effects in future calculations.

The principal strength of this approach lies in its modularity: the
kinematics are generated once and can subsequently be weighted
event by event with a weak-decay matrix element. This separation of kinematics and dynamics offers a computationally efficient pathway toward precision calculations of partial decay widths, differential spectra, and branching ratios. 

\bibliographystyle{unsrtnat}
\bibliography{phasespace_hypertriton}
\end{document}